\pdfoutput=1

\documentclass[11pt]{article}

\usepackage[final]{acl}
\usepackage{times}
\usepackage{comment,multirow,pgfplots}
\usepackage{makecell}

\usepackage{amssymb}
\usepackage[most]{tcolorbox}
\usepackage{caption}
\usepackage{algorithm}
\usepackage{algorithmic}
\usepackage{xspace}
\usepackage[table]{xcolor}

\usepackage[rgb]{xcolor} 
\definecolor{myblue}{rgb}{0.631, 0.682, 0.765}
\definecolor{myorange}{rgb}{0.749, 0.376, 0.0}
\newcommand{\BracketRank}{\mbox{\textbf{BracketRank}}\xspace}

 \usepackage{booktabs} 

\usepackage{graphicx}
\usepackage{listings}
\usepackage{subcaption}

\title{ BracketRank: Large Language Model Document Ranking via Reasoning-based Competitive Elimination
 }

\author{
  \textbf{Abdelrahman Abdallah, Mohammed Ali, Bhawna Piryani, Adam Jatowt} \\
  University of Innsbruck \\
  \texttt{\{abdelrahman.abdallah,adam.jatowt\}@uibk.ac.at}
}

\begin{document}
\maketitle

\begin{abstract}
Reasoning-intensive retrieval requires deep semantic inference beyond surface-level keyword matching, posing a challenge for current LLM-based rerankers limited by context constraints and order sensitivity. We propose \textbf{\BracketRank}, a framework that treats document reranking as a reasoning-driven competitive tournament. Our approach introduces three key innovations: (1) adaptive grouping based on model context limits, (2) reasoning-enhanced prompts that mandate step-by-step relevance explanations, and (3) a bracket-style elimination structure with winner and loser tracks. This design ensures robust document advancement while enabling parallel processing across competition stages. Evaluation on the BRIGHT reasoning benchmark shows that \BracketRank achieves \textbf{26.56 nDCG@10}, significantly outperforming state-of-the-art baselines including RankGPT-4 (17.0) and Rank-R1-14B (20.5).  On TREC datasets, BracketRank achieves 77.90 nDCG@5 on DL 19 and 75.85 nDCG@5 on DL 20, exceeding all baselines, establishing that explicit reasoning within competitive elimination is a powerful paradigm for complex, multi-step retrieval tasks.\footnote{\url{https://github.com/DataScienceUIBK/BracketRank}}
\end{abstract}

\section{Introduction}

Real-world information retrieval often demands more than surface-level matching between queries and documents. Finding documentation for a coding problem requires understanding function logic and syntax. Answering scientific questions necessitates connecting theoretical principles to specific phenomena. These \textit{reasoning-intensive} retrieval scenarios pose fundamental challenges that standard retrieval benchmarks fail to capture~\cite{su2024bright,abdallah2026tempo,abdallah2026mm,ali2026recor}. Existing benchmarks primarily consist of information-seeking queries where keyword or semantic matching suffices~\cite{bajaj2016ms,chen2017reading,thorne2018fever,lewis2020retrieval}. On the other hand, the BRIGHT benchmark~\cite{su2024bright} reveals that even state-of-the-art retrieval models achieve only 18.0 nDCG@10 on queries requiring intensive reasoning, which is dramatically lower than their performance on conventional benchmarks~\cite{bajaj2016ms,thakur2021beir}.

\begin{figure}[t]
    \centering
    \includegraphics[width=0.5\textwidth]{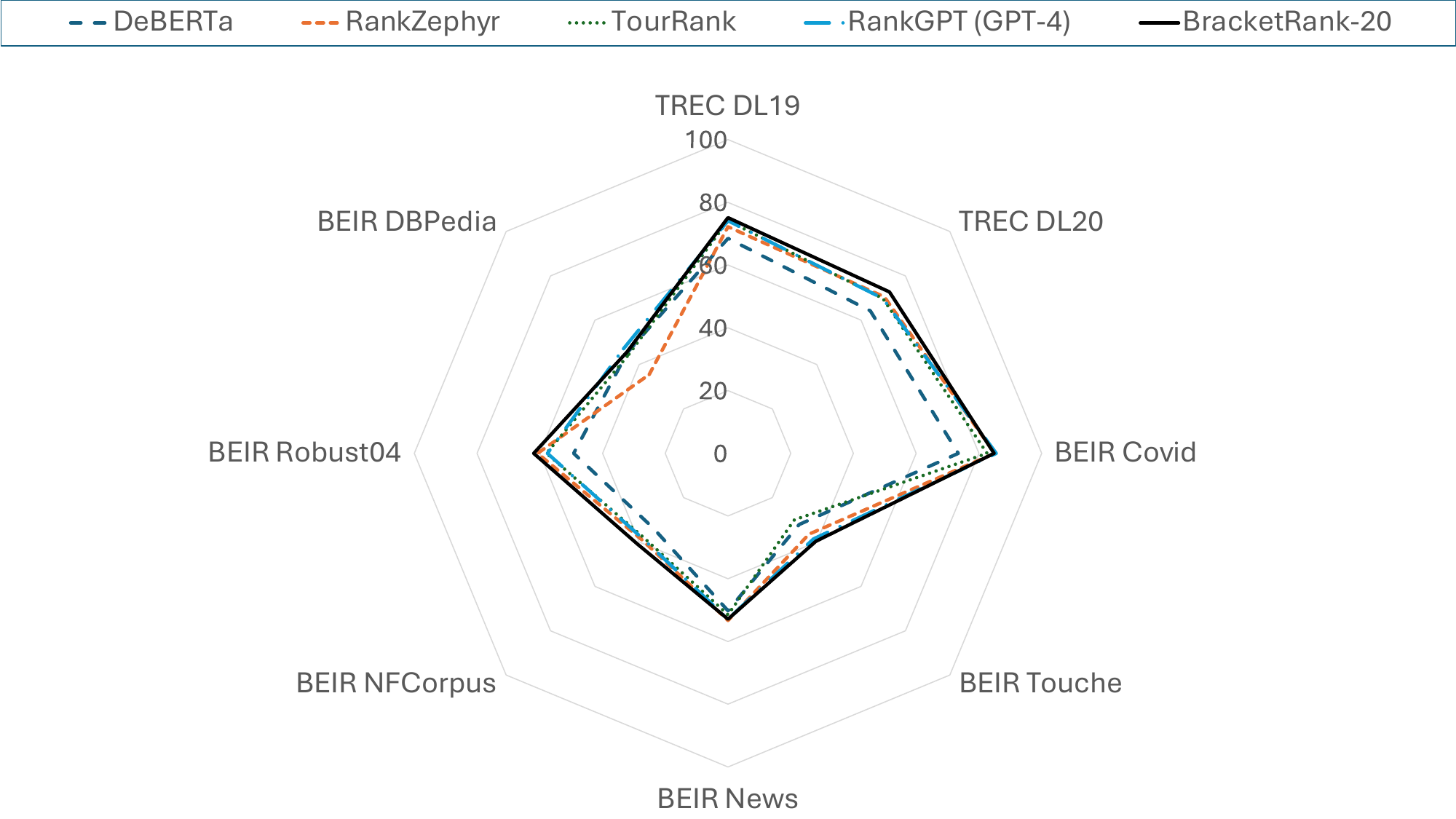}
    \caption{Radar chart comparing nDCG@5 performance of top reranking methods, including DeBERTa, RankZephyr, RankGPT (GPT-4), and \BracketRank-20 (GPT-4), across TREC DL20, TREC DL19 and BEIR datasets (Covid, NFCorpus, Touche, DBPedia, News, Robust04).}
    \label{fig:radar_chart_all_beir}
\end{figure}
\begin{figure*}[t]
    \centering
    \includegraphics[width=0.8\textwidth]{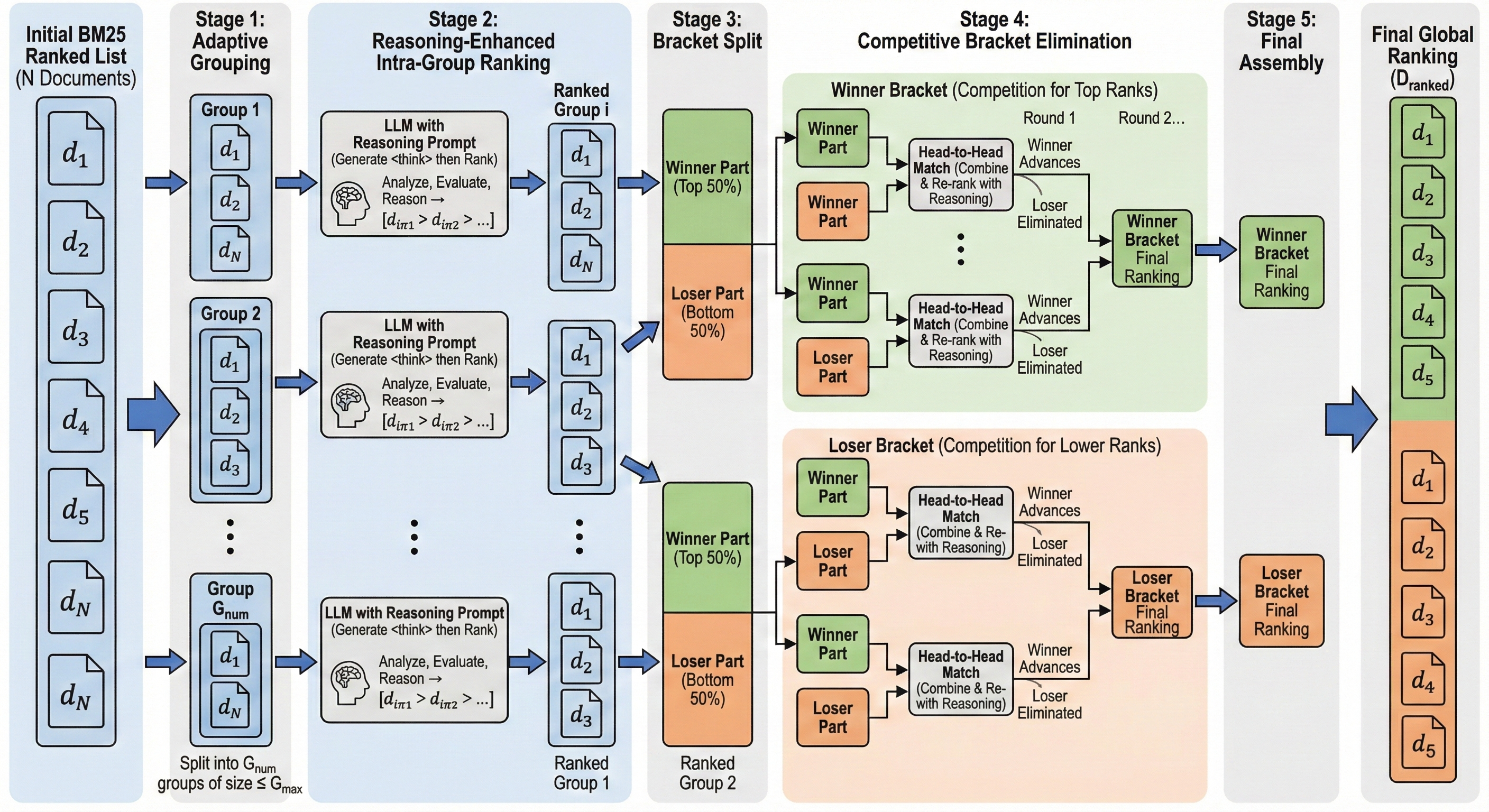} 
    \caption{Overview of the \BracketRank framework. The process consists of five stages: (1) adaptive grouping of initial retrievals; (2) intra-group ranking using LLMs with explicit reasoning prompts; (3) splitting ranked groups into winner and loser tracks; (4) parallel competitive bracket elimination where winners advance via head-to-head matches; and (5) assembly of the final global ranking. }
    \label{fig:bracketrank_main_overview}
\end{figure*}

Large Language Models (LLMs) offer a promising direction for reasoning-intensive retrieval through their demonstrated capabilities in complex reasoning tasks~\cite{wei2022chain}. LLM-based document ranking methods including pointwise~\cite{sachan2022improving,zhuang2023beyond}, pairwise~\cite{qin2023large}, and listwise~\cite{sun2023chatgpt,ma2023zero} approaches have shown strong zero-shot performance on traditional benchmarks. However, these methods face three key limitations that become especially problematic for reasoning-intensive tasks. First, context length constraints prevent processing many documents simultaneously, limiting the scope of comparative reasoning. Second, sequential decoding within listwise prompts creates bottlenecks that prevent parallelization. Third, ranking results depend heavily on initial document order, undermining the consistency needed for complex reasoning.

We propose \textbf{\BracketRank}, a reasoning-driven competitive elimination framework specifically designed for reasoning-intensive retrieval. Our key insight is that complex queries benefit from explicit, structured reasoning during document comparison---not just in final ranking decisions, but throughout a systematic competition process. \BracketRank combines three innovations: (1) adaptive grouping that optimizes group sizes based on LLM context limits, (2) reasoning-enhanced prompts that require step-by-step relevance explanations for each comparison, and (3) bracket-style elimination where documents compete through winner and loser brackets with explicit reasoning at each stage.

The competitive bracket structure provides two critical advantages for reasoning-intensive tasks. First, it enforces multiple rounds of deliberate comparison, allowing the LLM to reason about document relevance from different perspectives as documents face various opponents. Second, the winner/loser bracket design ensures documents receive fair evaluation regardless of initial positioning---particularly important when first-stage retrievers fail to identify relevant documents for complex queries.As shown in Figure~\ref{fig:radar_chart_all_beir}, \BracketRank achieves superior performance across diverse datasets, with particularly strong gains on reasoning-intensive benchmarks.


Our contributions are as follows: \textbf{(1)} We introduce \textbf{\BracketRank}, a competitive elimination framework that combines explicit reasoning requirements with systematic bracket competition, specifically designed for reasoning-intensive retrieval tasks. \textbf{(2)} We design reasoning-augmented prompts that improve ranking consistency by requiring step-by-step relevance explanations, enabling LLMs to perform deliberate comparative reasoning during document evaluation. \textbf{(3)} Extensive experiments demonstrate that \BracketRank achieves state-of-the-art performance on the BRIGHT benchmark for reasoning-intensive retrieval while maintaining competitive results on traditional benchmarks, establishing explicit reasoning as a key mechanism for complex retrieval tasks.

\section{Related Work}
Document ranking has evolved from BERT-based cross-encoders~\cite{abdallah2025dear,nogueira2019passage,devlin2018bert} to T5-based sequence models~\cite{nogueira2020document,raffel2020exploring}. LLM-based approaches fall into three categories: \textit{pointwise}~\cite{abdallah2025asrank,mozafari2025good,sachan2022improving,liang2022holistic}, \textit{pairwise}~\cite{mozafari2025good,qin2023large,pradeep2021expando}, and \textit{listwise}~\cite{mozafari2025good,sun2023chatgpt,ma2023zero}. Listwise methods like RankGPT~\cite{sun2023chatgpt} avoid quadratic pairwise costs but suffer from order sensitivity and sequential processing constraints. Tournament-inspired methods include ListT5~\cite{yoon2024listt5} (requires MS MARCO training) and TourRank~\cite{chen2025tourrank} (multi-round points accumulation). Our \BracketRank differs by using single-elimination brackets with explicit reasoning requirements for stable decisions. Recent work on reasoning for ranking~\cite{ji2024reasoningrank} focuses on explanation generation rather than consistency. See §~\ref{app:related} for detailed discussion.

\section{Method}
\label{sec:method}
\begin{figure*}[ht]
  \centering
  \begin{subfigure}[b]{0.45\textwidth}
    \centering
    \includegraphics[width=\textwidth]{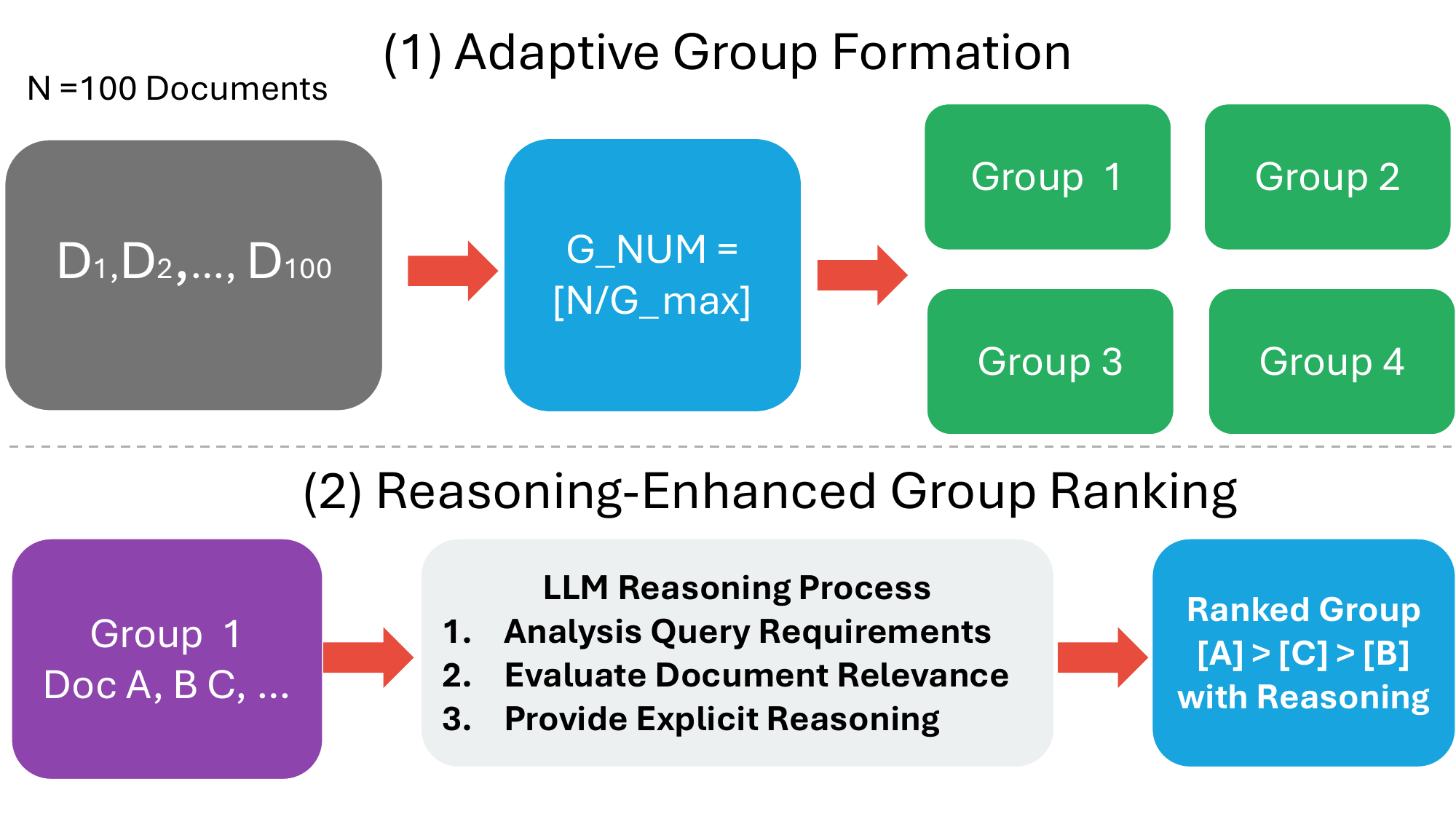}
    \caption{Adaptive group formation and reasoning-enhanced ranking within groups}
    \label{fig:bracketrank_process_appendix}
  \end{subfigure}
  \hfill
  \begin{subfigure}[b]{0.45\textwidth}
    \centering
    \includegraphics[width=\textwidth]{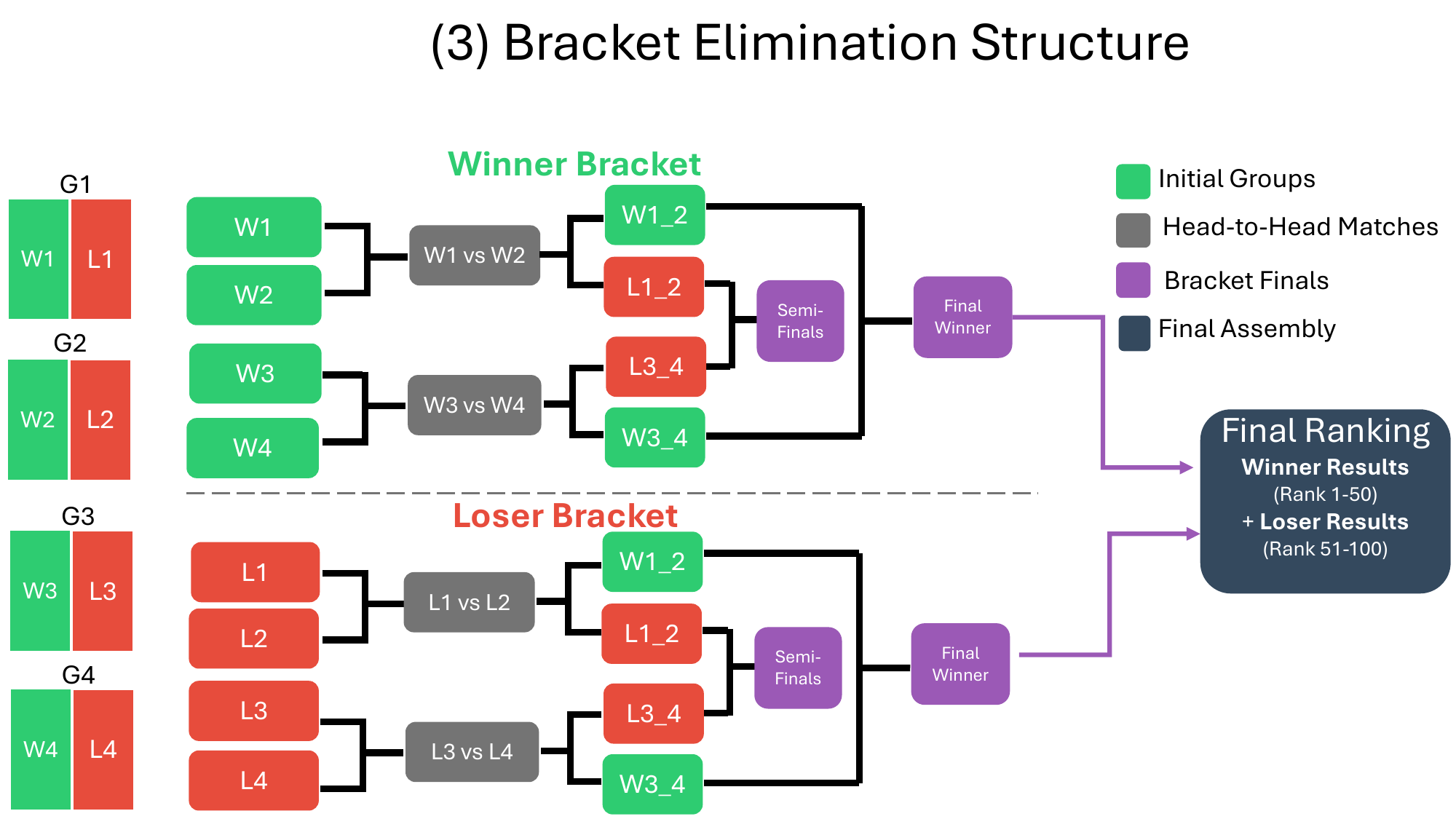}
    \caption{Competitive bracket elimination structure with winner and loser brackets}
    \label{fig:bracket_elimination_appendix}
  \end{subfigure}
  \caption{Detailed breakdown of \BracketRank methodology components. (a) illustrates the adaptive grouping based on context constraints. (b) details the specific document flow through the competitive elimination tracks.}
  \label{fig:bracketrank_overview}
\end{figure*}

In this section, we introduce \BracketRank, a reasoning-driven competitive elimination approach for zero-shot document ranking. Our method addresses the core limitations of existing listwise approaches through adaptive group formation and systematic bracket competition. We split ranked groups into winner and loser tracks and perform head-to-head matches by
\emph{combining two groups and re-ranking their documents} with the same listwise prompt.
Winners advance; losers are eliminated (single-elimination), yielding a global order after
$\lceil\log_2 G\rceil$ rounds. \BracketRank combines reasoning-enhanced prompts with tournament-style elimination to achieve robust rankings while maintaining efficiency. Figure \ref{fig:bracketrank_main_overview} illustrates the complete methodology.  Figure \ref{fig:bracketrank_overview} illustrates the complete methodology. We first explain how we form adaptive groups based on LLM context constraints. Then we describe the reasoning-enhanced ranking process within each group. Next, we detail the competitive bracket elimination that creates head-to-head competition between groups. Finally, we present the complete algorithm that integrates these components.

We first explain how we form adaptive groups based on LLM context constraints. Then we describe the reasoning-enhanced ranking process within each group. Next, we detail the competitive bracket elimination that creates head-to-head competition between groups. Finally, we present the complete algorithm that integrates these components.

\begin{algorithm}[t] \small
\caption{\BracketRank Algorithm: Adaptive group formation, reasoning-enhanced intra-group ranking, bracket split into winner and loser tracks, iterative elimination competition, and final ranking assembly}

\label{algorith_1}

\begin{algorithmic}[1]
\setlength{\itemsep}{-2pt} 
\STATE \textbf{Input:} Query $q$, candidate documents $D = \{d_1, \ldots, d_N\}$, max group size $G_{max}$
\STATE \textbf{Output:} Ranked document list $D_{ranked}$

\STATE // Adaptive group formation
\STATE $G_{num} \leftarrow \lceil N / G_{max} \rceil$
\STATE $Groups \leftarrow \text{SplitIntoGroups}(D, G_{num})$

\STATE // Reasoning-enhanced group ranking
\STATE $RankedGroups \leftarrow []$
\FOR{each $G_i$ in $Groups$}
    \STATE $R_i \leftarrow \text{ReasoningRank}(q, G_i)$
    \STATE $RankedGroups.\text{append}(R_i)$
\ENDFOR

\STATE // Create initial winner and loser brackets
\STATE $WinnerBracket \leftarrow []$, $LoserBracket \leftarrow []$
\FOR{each $R_i$ in $RankedGroups$}
    \STATE $W_i, L_i \leftarrow \text{SplitGroup}(R_i)$ // Top half vs bottom half
    \STATE $WinnerBracket.\text{append}(W_i)$
    \STATE $LoserBracket.\text{append}(L_i)$
\ENDFOR

\STATE // Iterative bracket elimination
\STATE $WinnerResult \leftarrow \text{RunIterativeBracket}(WinnerBracket, q)$
\STATE $LoserResult \leftarrow \text{RunIterativeBracket}(LoserBracket, q)$

\STATE // Combine final ranking
\STATE $D_{ranked} \leftarrow WinnerResult + LoserResult$
\STATE \textbf{return} $D_{ranked}$

\vspace{2mm}
\STATE \textbf{Function RunIterativeBracket}$(Bracket, q)$:
\WHILE{$|Bracket| > 1$}
    \STATE $NextRound \leftarrow []$
    \FOR{$i = 0$ to $|Bracket|-1$ step $2$}
        \IF{$i+1 < |Bracket|$}
            \STATE $Combined \leftarrow Bracket[i] + Bracket[i+1]$
            \STATE $Ranked \leftarrow \text{ReasoningRank}(q, Combined)$
            \STATE $Winner, Loser \leftarrow \text{SplitGroup}(Ranked)$
            \STATE $NextRound.\text{append}(Winner)$ // Only winners advance
        \ELSE
            \STATE $NextRound.\text{append}(Bracket[i])$ // Odd group advances
        \ENDIF
    \ENDFOR
    \STATE $Bracket \leftarrow NextRound$
\ENDWHILE
\STATE \textbf{return} $Bracket[0]$ // Final ranked list
\end{algorithmic}

\end{algorithm}



\subsection{Adaptive Group Formation}


Given $N$ candidate documents returned by the first-stage retriever, we determine the number and sizes of groups from the LLM's context budget. Let $G_{\max}$ denote the maximum number of documents that can fit into a single listwise prompt (respecting the model's token limit). We set the number of groups to
\begin{equation}
G_{\text{num}} \;=\; \left\lceil \frac{N}{G_{\max}} \right\rceil .
\label{eq:gnum}
\end{equation}

We partition the retriever-ranked list into $G_{\text{num}}$ \emph{contiguous} slices while keeping sizes as even as possible and never exceeding $G_{\max}$. Let
\[
s=\left\lfloor \frac{N}{G_{\text{num}}} \right\rfloor,\qquad
r = N \bmod G_{\text{num}}.
\]
We create $G_{\text{num}}$ groups in order: the first $r$ groups receive $s{+}1$ documents each and the remaining $G_{\text{num}}{-}r$ groups receive $s$ documents each. Denoting the size of group $i$ by $m_i$, we have
\[
m_i=\begin{cases}
s{+}1, & 1\le i \le r,\\[2pt]
s, & r{+}1\le i \le G_{\text{num}}.
\end{cases}
\]
This construction preserves the retriever order within every group and ensures that group-size differences are at most one. Because $G_{\text{num}}=\lceil N/G_{\max}\rceil$, it follows that $s{+}1 \le G_{\max}$, so no group exceeds the per-prompt limit. Concretely, if the groups are contiguous in the ranked list, their index ranges satisfy
\[
\begin{array}{l}
a_1=1,\quad b_i=a_i+m_i-1,\\
a_{i+1}=b_i+1 \quad (i=1,\dots,G_{\text{num}}{-}1).
\end{array}
\]

Given an LLM token budget $B$ for a single call, an average passage length $\bar{L}$ (in tokens), prompt overhead $T$ (instructions, delimiters), and per-document framing cost $H$ (e.g., identifiers and separators), a conservative choice is
\[
G_{\max} \;=\; \left\lfloor \frac{B - T}{\,\bar{L} + H\,} \right\rfloor .
\]
In our experiments we set $G_{\max}{=}20$ by estimating $(\bar{L},T,H)$ on the development split; \BracketRank is not sensitive to this choice within a reasonable range (see the group-size study in Fig.~\ref{fig:group_size_ablation}).

\subsection{Reasoning-Enhanced Group Ranking}
\begin{table}[h] \small

\centering
\resizebox{0.4\textwidth}{!}{
\begin{tabular}{c|c|c}
\toprule
\textbf{Stage} & \textbf{Number of Docs} & \textbf{Ranking Position} \\
\midrule
Initial Groups & $G \times D_g = N$ & Initial BM25 order \\
\midrule
Winner Groups & $G \times \lceil D_g/2 \rceil$ & Ranks 1 to $\sum W_i$ \\
\midrule
Loser Groups & $G \times \lfloor D_g/2 \rfloor$ & Ranks $\sum W_i + 1$ to $N$ \\
\midrule
Winner Champion & $W_{final}$ & Rank 1 \\
\midrule
Winner Runner-up & $W_{final}$ & Ranks 2 to $2 \times W_{final}$ \\
\midrule
$\vdots$ & $\vdots$ & $\vdots$ \\
\midrule
Loser Champion & $L_{final}$ & Rank $\sum W_i + 1$ \\
\midrule
Loser Final & $L_{final}$ & Ranks $N - L_{final} + 1$ to $N$ \\
\bottomrule
\end{tabular}
}

\caption{Document distribution and ranking structure in BracketRank. Groups are split into winner and loser brackets, with final rankings determined by bracket competition.}
\label{bracket_structure}
\end{table}
Each group undergoes independent ranking using reasoning-augmented prompts. This step transforms a relevance assessment into explicit reasoning about document quality relative to the query.

For a query $q$ and group $G_i = \{d_{i1}, d_{i2}, \ldots, d_{ik}\}$, 
the reasoning prompt structure is shown in Figure ~\ref{fig:bracketrank_prompt_appendix}. This reasoning component serves two purposes. First, it improves ranking by forcing the LLM to articulate its decision process. Second, it provides transparency into why certain documents rank higher than others.

Each group produces a ranked list $R_i = [d_{i\pi_1}, d_{i\pi_2}, \ldots, d_{i\pi_k}]$ where $\pi$ represents the ranking permutation. These group rankings form the input to the competitive elimination stage.

\subsection{Competitive Bracket Elimination}

The core innovation of \BracketRank lies in its competitive bracket structure. Rather than simply combining group results, we create systematic head-to-head competition between groups through winner and loser brackets.

\textbf{Winner and Loser Bracket Formation.} Each ranked group $R_i$ is split at the midpoint. The top half enters the winner bracket $W_i$ while the bottom half enters the loser bracket $L_i$. This creates two parallel competition tracks:

\begin{align}
W_i &= \{d_{i\pi_1}, d_{i\pi_2}, \ldots, d_{i\pi_{k/2}}\} \\
L_i &= \{d_{i\pi_{(k/2)+1}}, d_{i\pi_{(k/2)+2}}, \ldots, d_{i\pi_k}\}
\end{align}
\definecolor{color0}{rgb}{0.85,0.98,0.90} 
\definecolor{color1}{rgb}{0.70,0.95,0.80} 
\definecolor{color2}{rgb}{0.55,0.90,0.70} 
\definecolor{color3}{rgb}{0.35,0.85,0.60} 
\definecolor{color4}{rgb}{0.22,0.80,0.50} 

\begin{table*}[!t]
\centering
\scriptsize
\setlength\tabcolsep{2.5pt}
\resizebox{0.85\textwidth}{!}{
\begin{tabular}{l | ccccccc | cc | ccc | c}
\toprule
 & \multicolumn{7}{c|}{\textbf{StackExchange}} & \multicolumn{2}{c|}{\textbf{Coding}} & \multicolumn{3}{c|}{\textbf{Theorem-based}} & \\
\textbf{Method} & Bio. & Earth. & Econ. & Psy. & Rob. & Stack. & Sus. & Leet. & Pony & AoPS & TheoQ. & TheoT. & \textbf{Avg} \\
\midrule
BM25 & 18.2 & 27.9 & 16.4 & 13.4 & 10.9 & 16.3 & 16.1 & \cellcolor{color1}24.7 & 4.3 & 6.5 & 7.3 & 2.1 & 13.7 \\
\midrule
\multicolumn{14}{c}{\textit{Distilled / Fine-tuned Rerankers}} \\
\midrule
RankZephyr-7B & 21.9 & 23.7 & 14.4 & 10.3 & 7.6 & 13.7 & 16.6 & \cellcolor{color1}24.7 & 6.5 & 6.8 & 7.3 & 2.0 & 13.0 \\
Setwise-3B (SFT) & 22.0 & 18.8 & 10.4 & 11.5 & 9.1 & 5.8 & 16.7 & 9.9 & 5.7 & 4.0 & 3.8 & 3.4 & 10.1 \\
Setwise-7B (SFT) & \cellcolor{color1}28.7 & \cellcolor{color2}30.1 & 14.1 & \cellcolor{color2}23.9 & \cellcolor{color2}18.9 & 13.7 & 19.6 & \cellcolor{color2}20.7 & 7.1 & 7.0 & 8.2 & 8.2 & \cellcolor{color1}16.7 \\
Setwise-14B (SFT) & 22.0 & 29.3 & 15.4 & 23.0 & \cellcolor{color1}20.1 & \cellcolor{color1}15.7 & 20.3 & 19.4 & 6.2 & \cellcolor{color2}9.5 & \cellcolor{color2}9.7 & \cellcolor{color2}9.9 & \cellcolor{color1}16.7 \\
Rank-R1-3B (GRPO) & 18.4 & 17.1 & 13.7 & 16.9 & 9.0 & 10.0 & 16.5 & 11.1 & 4.7 & 3.5 & 3.2 & 5.9 & 10.8 \\
Rank-R1-7B (GRPO) & 26.0 & 28.5 & 17.2 & \cellcolor{color1}24.2 & \cellcolor{color2}19.1 & 10.4 & \cellcolor{color1}24.2 & 19.8 & 4.3 & 4.3 & 8.3 & \cellcolor{color1}10.9 & 16.4 \\
Rank-R1-14B (GRPO) & \cellcolor{color2}31.2 & \cellcolor{color1}38.5 & \cellcolor{color1}21.2 & \cellcolor{color2}26.4 & \cellcolor{color1}22.6 & \cellcolor{color2}18.9 & \cellcolor{color2}27.5 & \cellcolor{color1}20.2 & \cellcolor{color1}9.2 & \cellcolor{color1}9.7 & \cellcolor{color1}9.2 & \cellcolor{color1}11.9 & \cellcolor{color2}20.5 \\
\midrule
\multicolumn{14}{c}{\textit{Zero-Shot Rerankers}} \\
\midrule
Setwise-3B & 14.3 & 17.5 & 12.0 & 10.2 & 7.7 & 7.9 & 15.4 & 15.4 & 5.3 & 1.7 & 2.1 & 4.2 & 9.5 \\
Setwise-7B & 23.6 & 22.3 & 16.1 & 17.1 & 14.9 & 9.2 & 18.3 & 14.9 & 6.3 & 4.1 & 5.6 & \cellcolor{color1}10.4 & 13.6 \\
Setwise-14B & \cellcolor{color1}29.5 & \cellcolor{color2}32.2 & \cellcolor{color2}20.5 & \cellcolor{color1}24.8 & \cellcolor{color2}18.9 & 14.7 & \cellcolor{color1}23.6 & 18.7 & \cellcolor{color2}8.7 & 8.0 & 7.6 & 9.3 & \cellcolor{color1}18.0 \\
Rank-R1-3B & 13.7 & 17.3 & 11.9 & 15.2 & 10.0 & 6.6 & 17.8 & 7.7 & 3.7 & 4.0 & 2.5 & 6.0 & 9.7 \\
Rank-R1-7B & 26.8 & 24.8 & 17.9 & 22.1 & 17.4 & 10.3 & 21.1 & 15.6 & 4.4 & 3.3 & 5.9 & \cellcolor{color1}10.4 & 15.0 \\
Rank-R1-14B & \cellcolor{color2}30.1 & \cellcolor{color1}36.6 & \cellcolor{color1}22.1 & \cellcolor{color1}24.6 & \cellcolor{color1}21.7 & \cellcolor{color1}15.4 & \cellcolor{color2}25.0 & 17.0 & \cellcolor{color1}9.0 & \cellcolor{color1}9.1 & \cellcolor{color1}9.2 & \cellcolor{color2}11.6 & \cellcolor{color2}19.3 \\
RankGPT-4$^\dagger$ & \cellcolor{color2}33.8 & \cellcolor{color2}34.2 & 16.7 & \cellcolor{color2}27.0 & \cellcolor{color2}22.3 & \cellcolor{color2}27.7 & 11.1 & 3.4 & \cellcolor{color2}15.6 & 1.2 & 0.2 & 8.6 & 17.0 \\
\midrule
\BracketRank-20 (GPT-4) & \cellcolor{color4}\textbf{38.2}** & \cellcolor{color4}\textbf{45.2}*** & \cellcolor{color4}\textbf{22.7}* & \cellcolor{color4}\textbf{36.5}*** & \cellcolor{color4}\textbf{30.5}*** & \cellcolor{color4}\textbf{26.2}** & \cellcolor{color4}\textbf{33.8}*** & \cellcolor{color4}\textbf{16.9}*** & \cellcolor{color4}\textbf{28.7}** & \cellcolor{color4}\textbf{9.2} & \cellcolor{color4}\textbf{14.2}** & \cellcolor{color4}\textbf{16.6}*** & \cellcolor{color4}\textbf{26.6}*** \\
\bottomrule
\end{tabular}
}
\caption{BRIGHT nDCG@10 results for reasoning-intensive retrieval. All methods rerank BM25 top-100 documents. Statistical significance vs. best baseline: *p$<$0.05, **p$<$0.01, ***p$<$0.001. $^\wedge$Results from original paper using different BM25 system. \BracketRank achieves substantial improvements across all domains, with particularly strong gains on scientific reasoning (Biology, Earth Science, Psychology) and coding tasks (LeetCode, StackOverflow).}
\label{tab:bright}
\end{table*}

\definecolor{green0}{rgb}{0.85,0.98,0.90} 
\definecolor{green1}{rgb}{0.70,0.95,0.80} 
\definecolor{green2}{rgb}{0.55,0.90,0.70} 
\definecolor{green3}{rgb}{0.35,0.85,0.60} 
\definecolor{green4}{rgb}{0.22,0.80,0.50} 

\begin{table}[t!] \footnotesize
\centering
\resizebox{0.45\textwidth}{!}{%
\begin{tabular}{@{}l|l|cc|cc@{}}
\toprule
\multirow{2}{*}{\textbf{Methods}} & \multirow{2}{*}{\textbf{Model}} & \multicolumn{2}{c|}{\textbf{TREC DL 19}} & \multicolumn{2}{c}{\textbf{TREC DL 20}} \\
 &  &\textbf{NDCG@5} & \textbf{NDCG@10}  & \textbf{NDCG@5} & \textbf{NDCG@10}  \\
\midrule
BM25 & - &52.78 & 50.58  & 50.67 & 47.96  \\
\midrule
\multicolumn{6}{c}{\textbf{Supervised Methods}} \\
\midrule
monoBERT& BERT (340M) & 73.25 & 70.50 & 70.74 & 67.28  \\
monoT5 (220M)& T5 (220M) & \textbf{73.77} & 71.48  & 69.40 & 66.99  \\
monoT5& T5  (3B) & 73.74 & \cellcolor{green1}\textbf{71.83} &  \textbf{72.32} & \cellcolor{green0}\textbf{68.89}  \\
\midrule
\multicolumn{6}{c}{\textbf{Zero-Shot LLM Methods}} \\
\midrule
B-RG & GPT3.5 & 63.33 & 62.51 & 65.04 & 63.37  \\
PRP-Allpair& GPT3.5 & 70.43 & 68.18 & 69.75 & 66.40 \\
Setwise.heapsort (c=10)& GPT3.5 & 70.55 & 68.16  & 57.05 & 53.73 \\
Setwise.bubblesort (c=10)& GPT3.5 & 67.62 & 66.19 & 57.03 & 53.82 \\
RankGPT& GPT3.5 & 72.05 & 68.19 & 67.25 & 63.60 \\
RankGPT&  GPT-4 & \cellcolor{green1}75.98$^{**}$ & \cellcolor{green2}72.67$^{**}$  & 72.32 & \cellcolor{green1}69.48$^{*}$ \\
TourRank& GPT3.5 & \textbf{73.83}$^{*}$ & \textbf{71.63}$^{*}$  & \cellcolor{green1}\textbf{72.49}$^{*}$ & \cellcolor{green2}\textbf{69.56}$^{*}$ \\
TourRank &  GPT-4 & \cellcolor{green0}75.57$^{**}$ & \cellcolor{green3}74.13$^{***}$  & \cellcolor{green0}72.46$^{**}$ & \cellcolor{green3}69.79$^{**}$  \\
\midrule
\BracketRank$-10$ &  GPT-4   & \cellcolor{green4}\textbf{79.15}$^{***}$ & 65.66 &  \cellcolor{green3}\textbf{75.29}$^{***}$ & 64.04  \\
\BracketRank$-15$ &  GPT-4 & \cellcolor{green2}\textbf{77.53}$^{***}$ & \cellcolor{green0}71.65 &  \cellcolor{green2}\textbf{75.25}$^{***}$ & 67.56  \\
\BracketRank$-20$ &  GPT-4  & \cellcolor{green3}\textbf{77.90}$^{***}$ & \cellcolor{green4}\textbf{75.11}$^{***}$ &  \cellcolor{green4}\textbf{75.85}$^{***}$ & \cellcolor{green4}\textbf{72.75}$^{***}$  \\
\bottomrule
\end{tabular}
}
\caption{Performance comparison on TREC datasets. Best supervised and zero-shot methods shown in bold. Statistical significance vs. best baseline: $^{*}$p$<$0.05, $^{**}$p$<$0.01, $^{***}$p$<$0.001.}
\label{TREC}
\end{table}

\textbf{Head-to-Head Competition.} Groups compete in pairs through direct ranking. When two winner groups $W_i$ and $W_j$ compete, their documents are combined and re-ranked using the same reasoning-enhanced prompt. The winning group advances while the losing group is eliminated. This pairwise competition continues until each bracket produces a final ranking. 

\textbf{Final Ranking Assembly.} The complete ranking combines results from both brackets. Winner bracket results form the top portion of the final ranking, followed by loser bracket results. This guarantees that documents competing in the winner bracket receive higher positions than those in the loser bracket. Table~\ref{bracket_structure} shows how documents flow through the BracketRank structure. The adaptive grouping ensures balanced competition while the bracket elimination creates systematic advancement paths for all documents.


\subsection{The Complete BracketRank Algorithm}

Algorithm~\ref{algorith_1} operates in four phases. First, documents are divided into $G_{num} = \lceil N/G_{max} \rceil$ groups using contiguous splitting (lines 3-4). Second, each group undergoes independent reasoning-enhanced ranking that can be fully parallelized (lines 6-9). Third, ranked groups split at the midpoint into winner and loser brackets (lines 11-16). Fourth, iterative bracket elimination runs head-to-head competitions where groups combine and re-rank, with winners advancing until one group remains per bracket (lines 18-19). The algorithm supports two-level parallelisation. Initial group ranking processes all groups simultaneously. Bracket elimination processes all matches at the same round level in parallel. The computational complexity is $O(\log G_{num})$ for bracket elimination, making it significantly more efficient than $O(N^2)$ pairwise methods. Final ranking concatenates winner bracket results followed by loser bracket results, ensuring systematic relevance-based positioning. The competitive bracket structure (detailed in Algorithm~\ref{algorith_1}) ensures systematic advancement. We explored alternative tournament formats including double-elimination and round-robin (see §~\ref{app:BracketStructureComparison}), but single-elimination provides optimal performance-efficiency trade-offs.
\section{Experiments}


\subsection{Experimental Settings}


\paragraph{Datasets}

We evaluate BracketRank on benchmarks spanning both traditional and reasoning-intensive retrieval. 

\textbf{Reasoning-Intensive Retrieval.} \textbf{BRIGHT}~\cite{su2024bright} is the first benchmark specifically designed for reasoning-intensive retrieval, containing 1,384 real-world queries across 12 diverse datasets: Biology, Earth Science, Economics, Psychology, Robotics, StackOverflow, Sustainable Living (from StackExchange), LeetCode, Pony (coding domains), AoPS (math competitions), and TheoremQA-Theorem/Question. 

\textbf{Traditional Benchmarks.} \textbf{TREC DL 19 and DL 20}~\cite{craswell2020overview,craswell2021overview} contain 43 and 54 queries respectively for passage ranking evaluation. \textbf{BEIR}~\cite{thakur2021beir} provides heterogeneous zero-shot evaluation across eight diverse domains: Covid, NFCorpus, Touche, DBPedia, SciFact, Signal, News, and Robust04. \textbf{NovelEval-2306}~\cite{sun2023chatgpt} provides 21 queries from domains published after GPT-4's training cutoff, ensuring true zero-shot evaluation.

\paragraph{Metrics}

We re-rank the top-100 documents retrieved by BM25 using PySerini and Rankify implementation~\cite{lin2021pyserini,abdallah2025rankify}. Our evaluation uses NDCG@{1, 5, 10} as primary metrics, following standard practice in document ranking evaluation. 




\paragraph{Baselines}
We compare against supervised and zero-shot rerankers. \textbf{Supervised:} monoBERT~\cite{nogueira2019passage}, monoT5 (220M/3B)~\cite{nogueira2020document}, and Cohere Rerank-v2. \textbf{Zero-shot LLM:} B-RG~\cite{liang2022holistic}, PRP-Allpair~\cite{qin2023large}, Setwise~\cite{zhuang2023setwise}, RankGPT~\cite{sun2023chatgpt}, TourRank~\cite{chen2025tourrank}, and Rank-R1 (GRPO-trained)~\cite{zhuang2025rank}.



\subsection{Main Results}
\subsubsection{Performance on Reasoning-Intensive Retrieval}

Table~\ref{tab:bright} presents results on the BRIGHT benchmark for reasoning-intensive retrieval. \BracketRank achieves state-of-the-art performance with 26.6 average nDCG@10, substantially outperforming all baselines including the previous best Rank-R1-14B with GRPO training (20.5) and RankGPT-4 (17.0).  \BracketRank shows improvements across all 12 reasoning-intensive domains. The gains are particularly pronounced on scientific reasoning tasks: Biology (+7.0 points over Rank-R1-14B), Earth Science (+6.7 points), and Psychology (+10.1 points). 
On coding-related datasets, \BracketRank achieves 28.7 nDCG@10 on LeetCode (+8.5 points over RankGPT-4's cascaded approach) and 26.2 on StackOverflow. Understanding code requires analyzing function logic and syntax relationships precisely the type of multi-step reasoning that \BracketRank's explicit reasoning prompts facilitate.
Our Method provides substantial improvements on domains requiring multi-step inference, such as Biology and Psychology. A more detailed analysis of how explicit reasoning benefits specific BRIGHT categories is provided in Appendix \ref{sec:reasoning_analysis}.

\definecolor{color0}{rgb}{0.85,0.98,0.90} 
\definecolor{color1}{rgb}{0.70,0.95,0.80} 
\definecolor{color2}{rgb}{0.55,0.90,0.70} 
\definecolor{color3}{rgb}{0.35,0.85,0.60} 
\definecolor{color4}{rgb}{0.22,0.80,0.50} 

\begin{table}[!t]
\centering
\scriptsize
\setlength\tabcolsep{0.9pt}
\resizebox{0.5\textwidth}{!}{
\begin{tabular}{l | cccccccc | c }
\toprule
\textbf{Method} & Covid &  NFCorpus &  Touche &  DBPedia & SciFact &  Signal & News &  Robust04 & BEIR (Avg) \\
\midrule
BM25
 & 59.47 & 30.75 & \textbf{44.22} & 31.80 & 67.89 & 33.05 & 39.52 & 40.70 & 43.42
\\
\midrule
\multicolumn{10}{c}{\textbf{Supervised}}\\ 
\midrule
monoBERT (340M)
 & 70.01 & 36.88 & 31.75 & 41.87 & 71.36 & 31.44 & 44.62 & 49.35 & 47.16
\\
monoT5 (220M)  & 78.34 & 37.38 & 30.82 & 42.42 & 73.40 & 31.67 & 46.83 & 51.72 & 49.07
\\
monoT5 (3B)  & 80.71 & \cellcolor{color2}38.97 & 32.41 & \cellcolor{color0}44.45 &  \cellcolor{color1}76.57 & \cellcolor{color0}32.55 & 48.49 & \cellcolor{color0}56.71 & \cellcolor{color0}51.36
\\
Cohere Rerank-v2  & \cellcolor{color0}81.81 & 36.36 & 32.51 & 42.51 & 74.44 & 29.60 & 47.59 & 50.78 & 49.45
\\
ListT5 
(3B) & \cellcolor{color2}84.70 & 37.70 &  33.60&  \cellcolor{color3}46.20 & \cellcolor{color2}77.00& \cellcolor{color4}33.80    & \cellcolor{color4}53.20 &  57.80& \cellcolor{color2}53.00
\\
\midrule
\multicolumn{10}{c}{\textbf{Zero-Shot}}\\ 
\midrule
UPR (3B)
 & 68.11 & 35.04 & 19.69 & 30.91 & 72.69 & 31.91 & 43.11 & 42.43 & 42.99
\\
Promptagator++
 & 76.2 & 37.0 & \cellcolor{color1}38.1 & 43.4 & 73.1 & - & - & - & -
\\
RankGPT-3.5
 & 76.67 & 35.62 & 36.18 & 44.47 & 70.43 & 32.12 & 48.85 & 50.62 & 49.37
\\
RankGPT-4\textsuperscript{$\wedge$} & 85.51 & 38.47 & 38.57 & 47.12 & 74.95 & 34.40 & 52.89 & 57.55 & 53.68 \\
RankGPT (GPT-4) & 84.92 & 39.05 & 36.47 & 45.79 & 77.61 & 34.20 & 51.24 & 60.98 & 53.78 \\
TourRank (GPT-4)  &  \cellcolor{color1}82.59 & \cellcolor{color0}37.99 & 29.98 & \cellcolor{color1}44.64 & 72.17 & 30.83 & \cellcolor{color1}51.46 & \cellcolor{color2}57.87 & 50.94 \\
\midrule
\BracketRank (GPT-4) $-15$ 
 & 80.46 & \cellcolor{color3}40.04* &  \cellcolor{color0}35.94* & 43.95 & \cellcolor{color3}77.91*& 32.91 & \cellcolor{color0}49.85 &  \cellcolor{color3}59.57* & \cellcolor{color1}52.58*
\\
\BracketRank  (GPT-4) $- 20$ 
 & \cellcolor{color3}84.82 & \cellcolor{color4}\textbf{40.86**} & \cellcolor{color3}\textbf{39.67**} & \cellcolor{color2}45.58 & \cellcolor{color4}\textbf{78.03**} & \cellcolor{color2}33.49 & \cellcolor{color2}52.88* & \cellcolor{color4}\textbf{61.94**} & \cellcolor{color4}\textbf{54.66**}
\\
\bottomrule
\end{tabular} }
\caption{nDCG@10 performance on BEIR datasets (CovidQA, NFCorpus, Touche, DBPedia, SciFact, Signal, News, Robust04). Statistical significance vs. best baseline: *p$<$0.05, **p$<$0.01, ***p$<$0.001. RankGPT\textsuperscript{$\wedge$}: GPT\mbox{-}3.5 pre\mbox{-}rerank $\rightarrow$ GPT\mbox{-}4 on top\mbox{-}30 (original pipeline).
RankGPT (GPT\mbox{-}4): direct rerank on top\mbox{-}100 BM25 for fairness. }
\label{table:benchmark}
\end{table}

\subsubsection{Performance on Traditional and Novel Benchmarks}

\BracketRank achieves state-of-the-art results across traditional and novel benchmarks, outperforming supervised and zero-shot baselines. On \textbf{TREC DL 19 and DL 20}, \BracketRank-20 using GPT-4 reaches \textbf{77.90} and \textbf{75.85 nDCG@5}, respectively, significantly exceeding \textbf{RankGPT-4} ($75.98/72.32$) and \textbf{TourRank} ($75.57$) with statistical significance ($p < 0.001$). Notably, it surpasses supervised models like \textbf{monoT5 (3B)} in a purely zero-shot setting. On \textbf{BEIR}, \BracketRank-20 attains a leading average of \textbf{54.66 nDCG@10}, outperforming \textbf{RankGPT-4} ($53.68$) and \textbf{monoT5 (3B)} ($51.36$). It shows exceptional robustness in specialized domains, securing top scores in \textbf{NFCorpus} ($40.86$), \textbf{Touche} ($39.67$), \textbf{SciFact} ($78.03$), and \textbf{Robust04} ($61.94$). Furthermore, on \textbf{NovelEval-2306}, \BracketRank achieves an average \textbf{88.76 nDCG}, outperforming \textbf{RankGPT-4} ($87.88$) and confirming its effectiveness on content published after the model's training cutoff.

\begin{table}[!ht]
\centering
\small
\setlength\tabcolsep{1pt}
\resizebox{0.50\textwidth}{!}{
\begin{tabular}{l | cc | ccc | c}

\toprule
Method& prev. & Top-$K$ & nDCG@1 & nDCG@5 & nDCG@10 & Avg \\

\midrule
BM25
& - & -& 33.33 & 45.96 & 55.77 & 45.02\\
\midrule
monoBERT (340M)& BM25 & 100 & 78.57 & 70.65 & 77.27 & 75.50\\
monoT5 (220M)& BM25 & 100& 83.33 & 77.46 & 81.27 & 80.69\\
monoT5 (3B)& BM25 & 100& 83.33 & 78.38 & 84.62 & 82.11\\
\midrule

RankGPT-3.5& BM25 & 100& 76.19 & 74.15 & 75.71 & 75.35\\

RankGPT-4& RankGPT-3.5& 20& 85.71 & 87.49 & 90.45 & 87.88\\
\midrule
\BracketRank (GPT-4)    & BM25 & 100 & \textbf{86.42}   & \textbf{89.39} & \textbf{90.47} & \textbf{88.76}\\

\midrule
\end{tabular} }
\caption{Reranking results on NovelEval-2306. We compare BM25, monoT5, GPT baselines, and \BracketRank.}
\label{table:novel}
\end{table}


Table~\ref{table:novel} shows results on NovelEval-2306~\cite{sun2023chatgpt}, containing queries from domains published after GPT-4's training cutoff to ensure true zero-shot evaluation. \BracketRank achieves state-of-the-art performance with 88.76 average NDCG versus RankGPT-4's 87.88. Strong performance at NDCG@1 (86.42) and NDCG@5 (89.39) indicates the bracket competition structure effectively identifies relevant documents even without prior exposure to the content domain.


\begin{figure}[ht]
    \centering
    \includegraphics[width=0.45\textwidth]{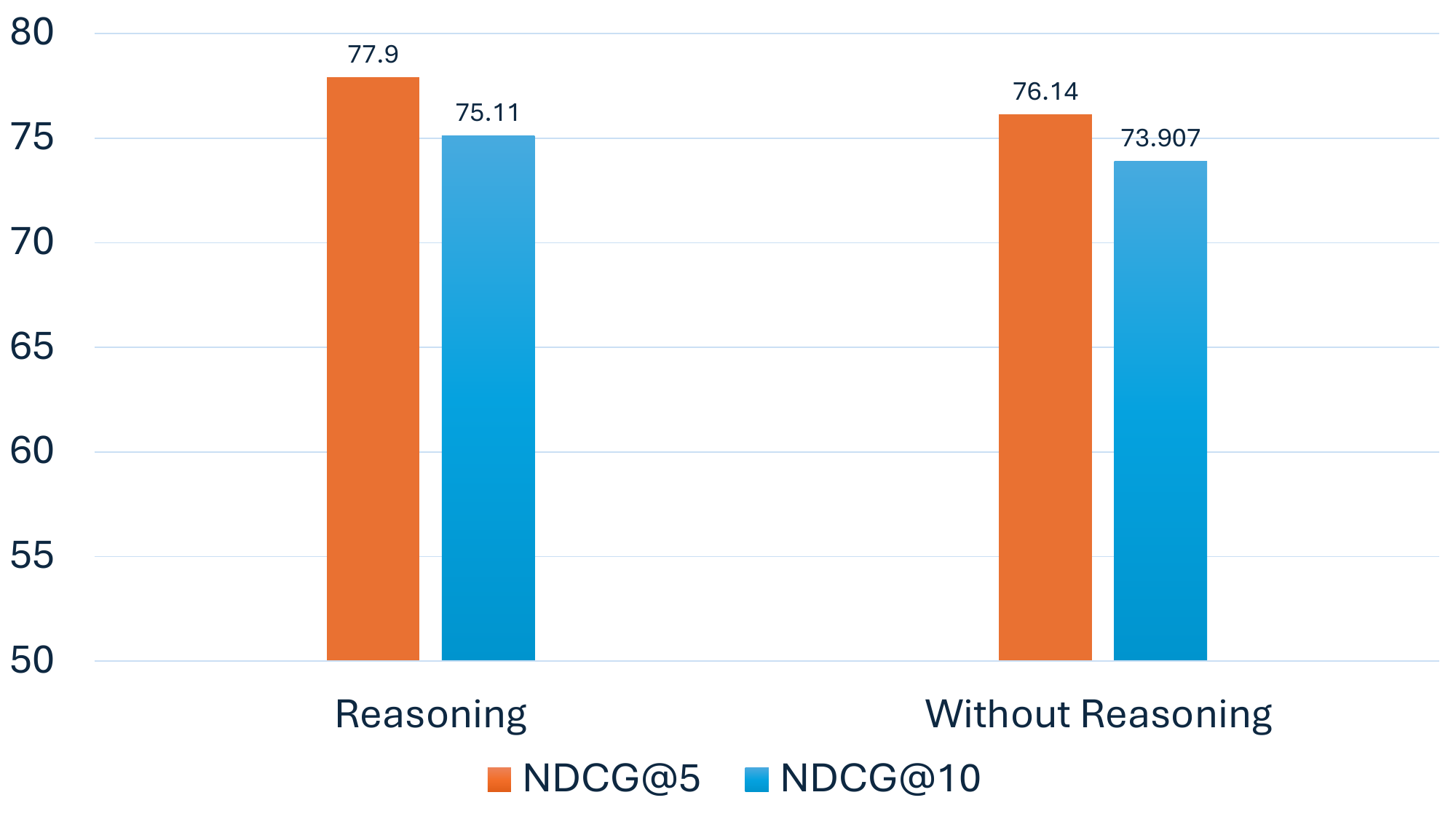}
    \caption{Ablation study showing the impact of reasoning-enhanced prompts on BracketRank performance. Reasoning requirements provide consistent improvements across both NDCG@5 and NDCG@10 metrics on TREC DL 19. In §~\ref{app:reason_vs_noreason}, we provide the reasoning versus bracket structure contributions on BEIR.}
    \label{fig:reasoning_ablation}
\end{figure}
\section{Additional Analysis}

\subsection{Impact of Reasoning Component}

To validate the effectiveness of reasoning-enhanced prompts, we conducted an ablation study on TREC DL 19 comparing BracketRank with and without explicit reasoning requirements. Figure~\ref{fig:reasoning_ablation} shows consistent improvements from the reasoning component. NDCG@5 increases from 76.14 to 77.90 (a gain of 1.76 points), while NDCG@10 improves from 73.91 to 75.11 (a gain of 1.20 points). The reasoning requirements force LLMs to articulate their relevance judgments explicitly, reducing inconsistencies that occur with implicit decisions. This structured approach helps models consider multiple relevance signals systematically and creates more stable rankings when documents encounter. The reasoning requirements force LLMs to articulate their relevance judgments explicitly, reducing inconsistencies that occur with implicit decisions. This structured approach helps models consider multiple relevance signals systematically and creates more stable rankings. We provide additional analysis of reasoning versus bracket structure contributions on BEIR datasets in §~\ref{app:reason_vs_noreason}.

\begin{table}[!t]
\centering
\small
\setlength\tabcolsep{2pt}
\resizebox{0.4\textwidth}{!}{
\begin{tabular}{l |l|ccc c}

\toprule
Method &  Model & @1 & @5 & @10 & Avg \\

\midrule

BM25
& -& 54.26 & 52.78 & 50.58 & 52.54
\\







\midrule
\BracketRank  & GPT-4       
& 79.07 & 77.90 & 75.11 & 77.36\\

\BracketRank & Qwen2.5-7B    
 & 76.36  & 72.32 & 67.58 & 72.09
\\

\BracketRank   &Llama-3.1-8B
 &  69.77 &66.42&  63.84 &  66.68
\\

\BracketRank    & Gemini-2.5-flash &
 78.68 & 74.88 &  71.90 &  75.15 \\

\BracketRank    &Gemini-2.5-pro  &
 79.46 & 74.41 &  72.86 &  75.58

\\

\bottomrule
\end{tabular}}

\caption{Results of NDGC for different LLMs on DL-19.}

\label{table:more-llm}
\end{table}



\begin{table}[ht!]
\centering
\small
\resizebox{0.45\textwidth}{!}{
\begin{tabular}{l|l|cc}
\toprule
\textbf{Model} & \textbf{Method} & \textbf{NDCG@5} & \textbf{NDCG@10} \\
\midrule
\multirow{2}{*}{Qwen2.5-7B} & TourRank & 67.24 & 64.48 \\
& \BracketRank & \textbf{72.32} (+5.08) & \textbf{67.58} (+3.10) \\
\midrule
\multirow{2}{*}{Llama-3.1-8B} & TourRank & 65.09 & 62.35 \\
& \BracketRank & \textbf{66.42} (+1.33) & \textbf{63.84} (+1.49) \\
\bottomrule
\end{tabular}
}
\caption{Cross-LLM comparison on TREC DL19 showing \BracketRank consistently outperforms TourRank across different model backends.}
\label{tab:cross_llm}
\end{table}

\subsection{Cross-LLM Robustness Analysis}

We evaluate \BracketRank across proprietary and open-source LLMs to assess framework generalizability. Table~\ref{table:more-llm} shows performance ranges from 66.68 (Llama-3.1-8B) to 77.36 (GPT-4) on TREC DL-19, demonstrating that the competitive elimination framework leverages fundamental reasoning abilities rather than model-specific features. Even the open-source Qwen2.5-7B achieves 72.09, making \BracketRank viable for resource-constrained deployments. Direct comparison with TourRank (Table~\ref{tab:cross_llm}) confirms advantages persist across backends: +5.08 NDCG@5 with Qwen2.5-7B and +1.33 with Llama-3.1-8B. More capable LLMs better exploit the bracket structure, but smaller models still benefit from systematic elimination.




\begin{figure}[ht]
    \centering
    \includegraphics[width=0.5\textwidth]{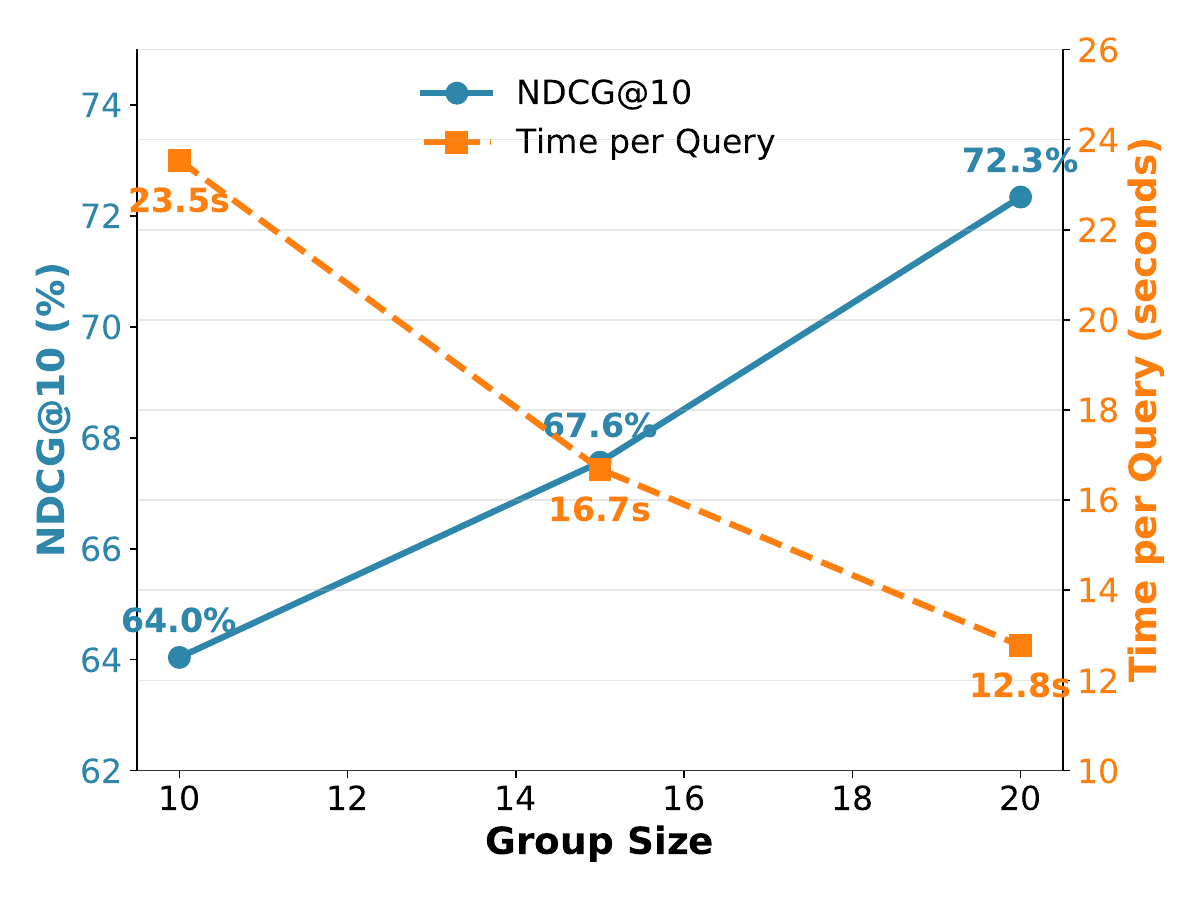}
\caption{ Group Size Impact on Per-Query Performance and Efficiency}
    \label{fig:group_size_ablation}
\end{figure}

\subsection{Group Size and Complexity}

Larger group sizes improve both performance and efficiency (Figure~\ref{fig:group_size_ablation}). BracketRank-20 achieves 72.34\% NDCG@10 versus 64.04\% for BracketRank-10, while reducing processing time from 23.54 to 12.77 seconds per query (45.7\% reduction). Larger groups enable more comprehensive intra-group comparisons and fewer bracket elimination rounds. \BracketRank requires $O(\log G_{num})$ rounds with approximately $N(1 + \log G_{num})$ document inputs. For $N=100$ and $G_{max}=20$, this yields ~332 inputs—more efficient than pairwise methods while maintaining superior quality. Detailed complexity analysis is in §~\ref{app:Theoretical_compleity}.

\begin{table}[ht!]
\centering
\small
\resizebox{0.5\textwidth}{!}{
\begin{tabular}{l|l|cc|c}
\toprule
\textbf{Method} & \textbf{Retriever} & \textbf{DL 19} & \textbf{DL 20} & \textbf{Avg} \\
& & \textbf{NDCG@10} & \textbf{NDCG@10} & \\
\midrule
Contriever & - & 62.02 & 63.42 & 62.72 \\
\midrule
RankGPT & Contriever & 69.70 & 68.47 & 69.09 \\
TourRank-2 & Contriever & 69.12 & 71.89 & 70.51 \\
TourRank-10 & Contriever & 70.77 & 73.19 & 71.98 \\
\textbf{BracketRank-20} & Contriever & \textbf{72.32} & \textbf{74.64} & \textbf{73.48} \\
\bottomrule
\end{tabular}
}
\caption{Performance with dense retriever (Contriever) on TREC datasets, demonstrating robustness across different retrieval paradigms.}
\label{tab:contriever}
\end{table}

\subsection{Retriever Robustness}

While our main setup uses BM25, we also test with the dense retriever Contriever~\cite{izacard2021contriever}. Table~\ref{tab:contriever} shows that \BracketRank retains its edge under dense retrieval: it averages 73.48 NDCG@10 on TREC DL, vs.\ 71.98 for TourRank-10 and 69.09 for RankGPT. The +4.39 vs.\ RankGPT and +1.50 vs.\ TourRank gains indicate the tournament design is not tied to a sparse retriever.

\vspace{-3mm}
\subsection{Ablation and Efficiency Analysis}

Ablations on DL 20 (Table~\ref{tab:component_ablation}) confirm that combining intra-group listwise ranking and bracketed elimination is essential, yielding \textbf{+4.62} and \textbf{+8.73 nDCG@5} gains over group concatenation and pairwise variants, respectively. Single-elimination provides the optimal performance-efficiency trade-off compared to double-elimination or round-robin formats. Regarding computational costs (Table~\ref{tab:api_efficiency}), \BracketRank-20 requires \textbf{13 API calls} (360 total documents), providing greater comparison depth than RankGPT (9 calls, 200 documents) while using \textbf{7.7$\times$ fewer calls} than TourRank-10 to achieve superior quality. See §~\ref{app:ComputationalPractice} for detailed analysis.



\begin{table}[t!]
\centering
\small
\resizebox{0.5\textwidth}{!}{
\begin{tabular}{l|cc|l}
\toprule
\textbf{Approach} & \textbf{NDCG@5} & \textbf{NDCG@10} & \textbf{Description} \\
\midrule
Listwise Only & 71.23 & 68.45 & Groups ranked, then concatenated \\
Pairwise Groups & 67.12 & 64.88 & Pairwise group comparisons \\
\midrule
\textbf{BracketRank (Full)} & \textbf{75.85} & \textbf{72.75} & Reasoning + bracket elimination \\
\bottomrule
\end{tabular}
}
\caption{Component ablation on TREC DL20 isolating contributions of bracket elimination structure vs. reasoning-enhanced group ranking.}
\label{tab:component_ablation}
\end{table}

\begin{figure}[t]
    \centering
    \includegraphics[width=0.5\textwidth]{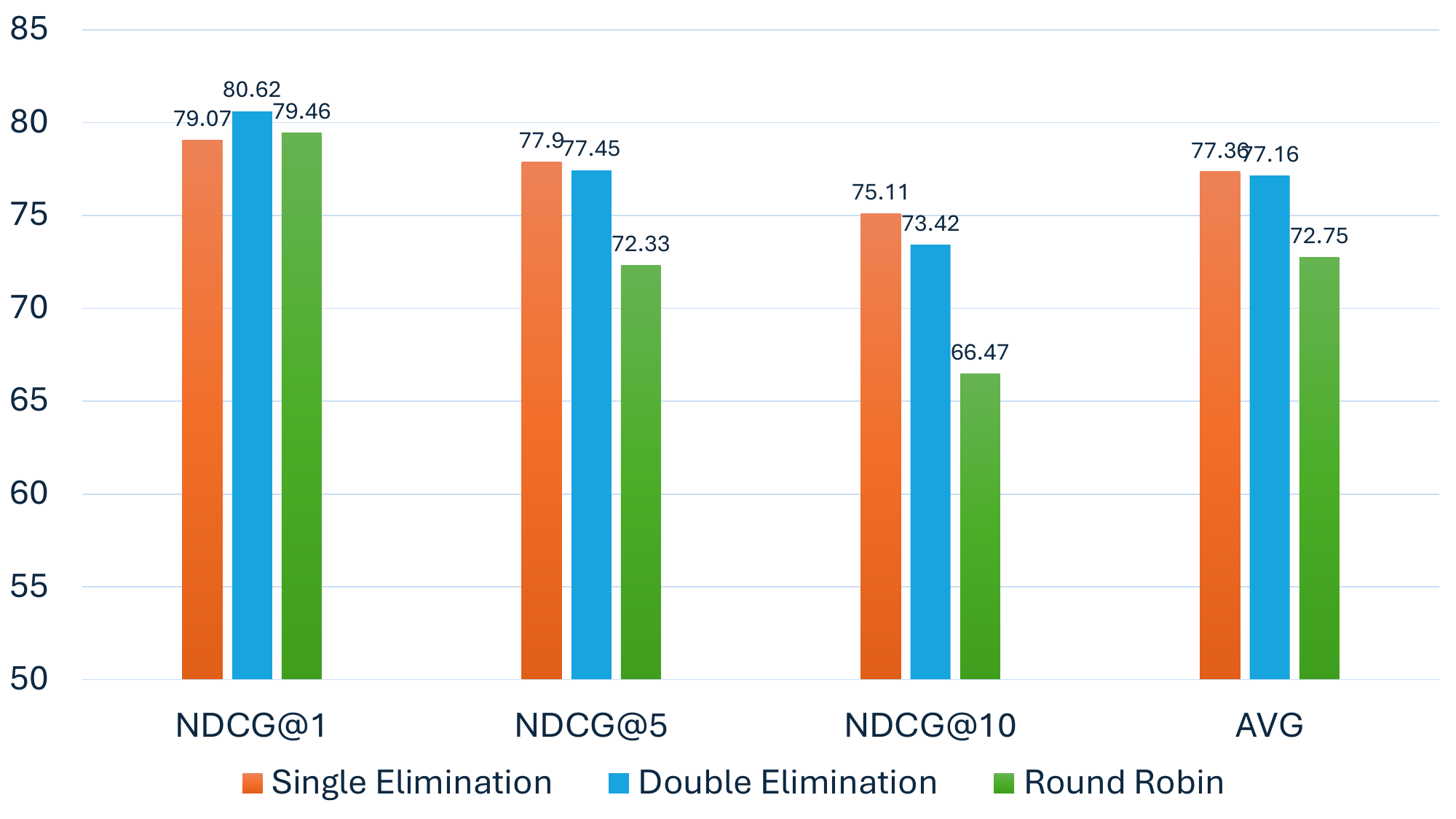}
\caption{Comparison of bracket elimination structures on TREC DL 19. Single elimination achieves the best overall performance across all NDCG}
    \label{fig:stra}
\end{figure}

\subsection{Bracket Structure Comparison}
\label{app:BracketStructureComparison}
We evaluate three different competitive structures to validate our single elimination design choice. The structures are defined as follows:

\textbf{Single Elimination}: Our current approach where groups compete once, with winners and losers split into separate brackets that run parallel elimination tournaments. \textbf{Double Elimination}: Groups receive two chances before elimination. Losers from the winner bracket drop to a loser bracket, and only groups that lose twice are eliminated completely. \textbf{Round Robin}: Every group competes against every other group in exhaustive pairwise comparisons. Final rankings are determined by win-loss records and average document positions across all matches.

Figure~\ref{fig:stra} compares performance across these bracket strategies on TREC DL 19. Single elimination achieves the best overall performance (77.36 average NDCG), demonstrating optimal balance across all ranking cutoffs. Double elimination reaches the highest precision at NDCG@1 (80.62), suggesting that second chances help identify the most relevant documents. However, performance drops significantly at higher cutoffs (NDCG@5: 77.45, NDCG@10: 73.42), indicating that additional complexity introduces ranking inconsistencies. Round robin shows substantial performance degradation, particularly at NDCG@10 (66.47). While exhaustive comparison seems theoretically superior, the excessive pairwise competition creates noise that degrades ranking quality. The $O(n^2)$ computational overhead also makes this approach impractical for larger document sets.

\section{Efficiency-Effectiveness Trade-off}

\begin{figure}[htbp]
    \centering
    \includegraphics[width=0.4\textwidth]{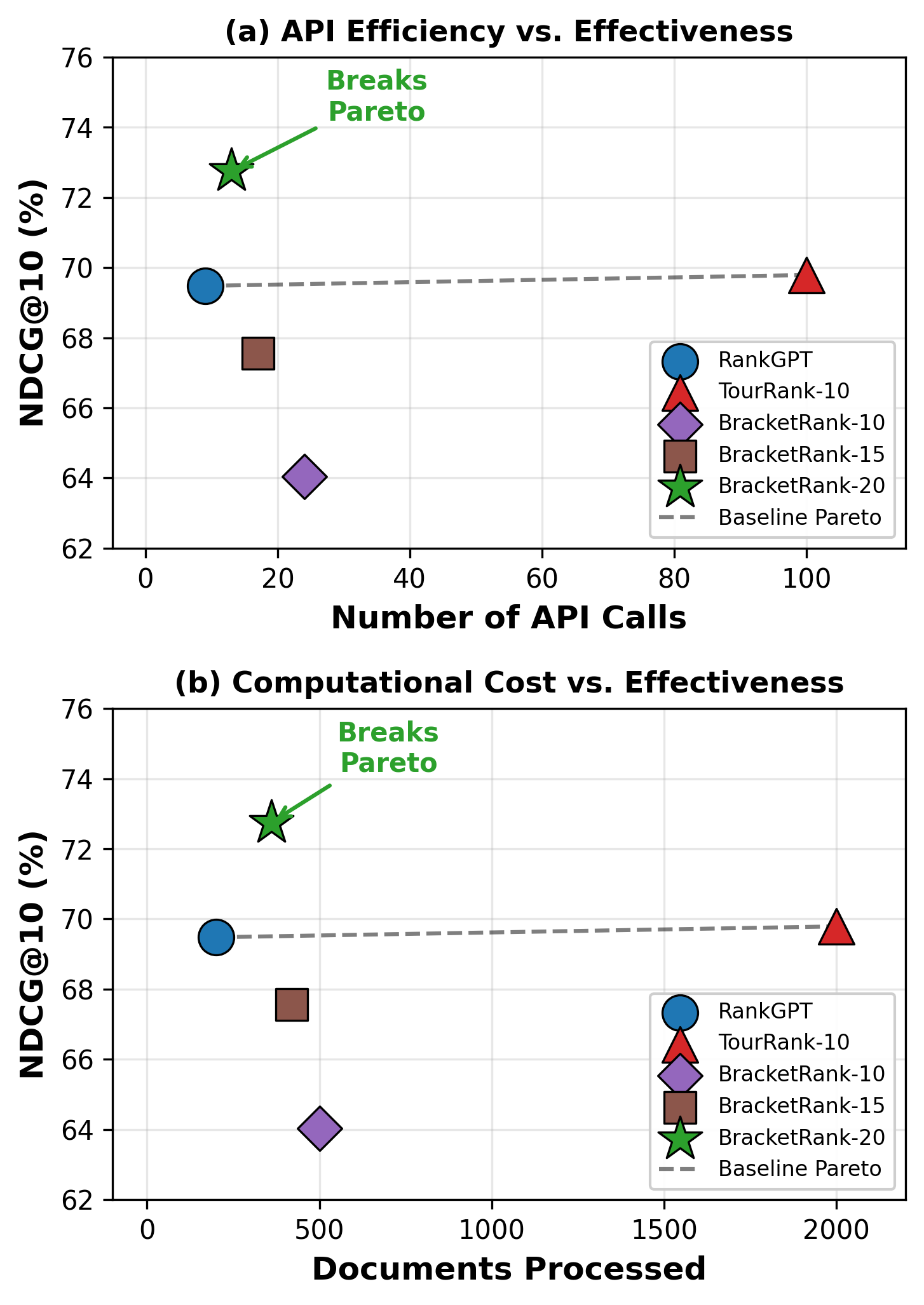}
    \caption{Efficiency-Effectiveness Pareto analysis on TREC DL20. }
    \label{fig:pareto_vertical}
\end{figure}

To evaluate the practical utility of \BracketRank, we analyze the trade-off between ranking effectiveness (NDCG@10) and computational overhead. Figure \ref{fig:pareto_vertical} illustrates this balance across two dimensions: \textbf{API Efficiency} (number of requests) and \textbf{Computational Cost} (total documents processed).

As shown in Figure \ref{fig:pareto_vertical}(a), \BracketRank-20 effectively \textbf{breaks the existing Pareto front} defined by current state-of-the-art methods. While RankGPT is efficient, its effectiveness is limited by its sliding-window approach. Conversely, TourRank-10 improves effectiveness at an $11\times$ higher API cost. \BracketRank-20 achieves superior performance (72.75\% NDCG@10) while requiring only 13 API calls—an 87\% reduction in cost compared to TourRank-10. This confirms that our reasoning-driven competitive elimination provides a superior "exchange rate" of 0.82 NDCG points per additional API call relative to RankGPT. For a detailed theoretical derivation and exchange rate analysis, see \textbf{Appendix \ref{app:efficiency_analysis}}.


\section{Conclusion}
\label{sec:conclusion}

We introduced \BracketRank, a reasoning-driven competitive elimination framework for LLM-based document ranking. Our method combines adaptive grouping, reasoning-enhanced prompts, and systematic bracket competition to address key limitations in existing listwise approaches. \BracketRank achieves state-of-the-art performance with 77.90 NDCG@5 on TREC DL 19 and 54.66 average NDCG@10 across BEIR datasets.
\section*{Acknowledgments}
The computational results presented here have been achieved (in part) using the LEO HPC infrastructure of the University of Innsbruck.



\section*{Limitations}
\label{sec:limitations}

While \BracketRank achieves state-of-the-art performance across reasoning-intensive benchmarks, its deployment involves standard trade-offs common to advanced LLM-based reranking frameworks.

First, the method is subject to \textbf{inherent model constraints} such as context window limits and API latency. While our adaptive grouping strategy ensures compatibility with diverse model architectures, extremely long document collections may require standard truncation or multi-stage processing. These constraints are characteristic of all LLM-based listwise approaches and are typically mitigated by the continuous advancement of underlying model capacities and hardware acceleration.


Second, the \textbf{reasoning-enhanced prompts} are optimized for current instruction-following LLMs. While we observed consistent gains across various model families, the depth of generated explanations naturally scales with the size of the backbone model. We view these constraints as opportunities for future research into distilling bracket-ranking logic into smaller, task-specific models to further reduce operational overhead while maintaining reasoning depth.

\bibliography{custom}
\appendix



\section{Results}

\subsection{Theoretical Analysis of Efficiency}
\label{app:Theoretical_compleity}
Table \ref{bracketrank_complexity} shows the theoretical time complexity and document usage comparison between \BracketRank and existing methods. The analysis is based on our adaptive grouping strategy with maximum group size $G_{max}$ and the bracket elimination structure. From Table \ref{bracketrank_complexity}, we can observe several key insights about computational efficiency and document processing requirements.

\begin{table}[th]

\centering
\resizebox{0.4\textwidth}{!}{%
\begin{tabular}{ccc}
\toprule
\textbf{Methods} & \textbf{Time Complexity} & \textbf{No. of Docs to LLMs} \\ 
\midrule
PointWise & $O(1)$ & $N$ \\
\midrule
PRP-Allpair & $O(1)$ & $N^2-N$ \\
\midrule
Setwise.bubblesort & $\approx O(\frac{1}{9}k \cdot N)$ & $\approx \frac{10}{9}k \cdot N$ \\
\midrule
Setwise.heapsort & $\approx O(k \cdot \log_{10} N)$ & $\approx 10k \cdot \log_{10} N$ \\
\midrule
RankGPT & $\approx O(\frac{1}{10} \cdot N)$ & $\approx 2N$ \\
\midrule
TourRank-$r$ & $O(K-1)$ & $\approx 2r \cdot N$ \\
\midrule
\textbf{\BracketRank} & $O(\log G_{num})$ & $\approx N(1 + \log G_{num})$ \\
\bottomrule
\end{tabular}%
}

\caption{Theoretical time complexity and document input requirements for different ranking methods. $N$ is the number of candidate documents, $G_{max}$ is the maximum group size constraint in \BracketRank (default: 20), $G_{num} = \lceil N/G_{max} \rceil$ is the number of groups, $k=10$ is the number of documents compared per prompt in Setwise methods, and $r$ is the number of tournament rounds in TourRank.}

\label{bracketrank_complexity}
\end{table}

\textbf{Time Complexity Analysis.} \BracketRank achieves $O(\log G_{num})$ time complexity through its parallel bracket elimination structure. Since $G_{num} = \lceil N/G_{max} \rceil$ and our default $G_{max} = 20$, this translates to $O(\log(N/20)) \approx O(\log N)$ for practical datasets. 

\textbf{Document Processing Efficiency.} \BracketRank requires approximately $N(1 + \log G_{num})$ document inputs to LLMs. This consists of the initial group ranking phase (contributing $N$ documents) and the bracket elimination phases (contributing $N \log G_{num}$ documents as documents get re-ranked in each elimination round). For typical datasets with $N=100$ and $G_{max}=20$, this results in roughly $100(1 + \log 5) \approx 332$ document inputs, which is more efficient than pairwise methods but involves more processing than simple listwise approaches.

\subsection{Disentangling Gains: Bracket vs.\ Reasoning}
\label{app:reason_vs_noreason}
We separate the impact of the tournament structure from the reasoning prompt on BEIR. Table~\ref{tab:beir_reasoning} shows that competitive elimination alone lifts Avg NDCG@10 from 50.94 (TourRank) to 53.79 (+2.85), with large gains on Touche (+4.61), SciFact (+6.90), and Robust04 (+4.68). Adding the reasoning step yields a further +0.87 to 54.66, indicating most of the improvement comes from the bracketed competition, with reasoning providing a consistent secondary boost.

\begin{table*}[t!]
\centering
\scriptsize
\setlength\tabcolsep{0.9pt}
\resizebox{0.85\textwidth}{!}{
\begin{tabular}{l | cccccccc | c }
\toprule
\textbf{Method} & Covid & NFCorpus & Touche & DBPedia & SciFact & Signal & News & Robust04 & \textbf{Avg} \\
\midrule
TourRank (GPT-4) & 82.59 & 37.99 & 29.98 & 44.64 & 72.17 & 30.83 & 51.46 & 57.87 & 50.94 \\
\midrule
\BracketRank-20 (no reasoning) & 84.66 & 39.79 & 34.59 & 45.42 & 79.07 & 33.28 & 50.93 & 62.55 & 53.79 \\
\BracketRank-20 (with reasoning) & \textbf{84.82} & \textbf{40.86} & \textbf{39.67} & \textbf{45.58} & \textbf{78.03} & \textbf{33.49} & \textbf{52.88} & \textbf{61.94} & \textbf{54.66} \\
\bottomrule
\end{tabular}}
\caption{BEIR results (NDCG@10) showing BracketRank performance with and without reasoning component, demonstrating that competitive elimination provides substantial gains independent of prompt design.}
\label{tab:beir_reasoning}
\end{table*}
\subsection{Computational Efficiency in Practice}
\label{app:ComputationalPractice}
Table~\ref{tab:api_efficiency} presents a comprehensive comparison of computational requirements across methods. \BracketRank-20 requires 13 API calls per query, processing 360 total documents with an average of 27.7 documents per call. This represents a middle ground between RankGPT's minimal 9 calls and TourRank-10's extensive 100 calls. While RankGPT appears more efficient with fewer API calls (9 vs 13), it processes only 200 total documents compared to \BracketRank's 360, limiting the depth of comparison and contributing to its lower ranking quality.

\begin{table}[t!]
\centering
\small
\resizebox{0.48\textwidth}{!}{
\begin{tabular}{l|ccc}
\toprule
\textbf{Method} & \textbf{API Calls} & \textbf{Total Docs} & \textbf{Docs per Call} \\
& & \textbf{Processed} & \\
\midrule
RankGPT & 9 & 200 & 22.2 \\
\textbf{BracketRank-20} & \textbf{13} & \textbf{360} & \textbf{27.7} \\
TourRank-10 & 100 & 2000 & 20.0 \\
PRP-Allpair & 9900 & 9900 & 1.0 \\
\bottomrule
\end{tabular}
}
\caption{API usage comparison under matched conditions (100 documents, TREC DL20). \BracketRank achieves superior performance with moderate computational requirements, balancing efficiency and ranking quality.}
\label{tab:api_efficiency}
\end{table}

\section{Related Work}
\label{app:related}
Document ranking has advanced with pretrained models like BERT~\cite{devlin2018bert} and T5~\cite{raffel2020exploring}. Nogueira and Cho~\cite{nogueira2019passage} developed monoBERT and duoBERT for quality-speed balance. ~\citet{nogueira2020document} adapted T5 using sequence-to-sequence frameworks. ~\citet{zhuang2023rankt5} introduced RankT5 with ranking-specific losses.

\textbf{Pointwise Methods.} Query Generation~\cite{sachan2022improving} rescores passages by computing question probability. Binary Relevance Generation~\cite{liang2022holistic} uses "Yes/No" predictions. Rating Scale methods~\cite{zhuang2023beyond} incorporate fine-grained relevance labels. ~\citet{guo2024generating} propose multi-perspective evaluation criteria.

\textbf{Pairwise Methods.} ~\citet{pradeep2021expando} design T5-based pairwise components. ~\citet{qin2023large} introduce Pairwise Ranking Prompting (PRP). These approaches capture document relationships but require quadratic comparisons.

\textbf{Listwise Methods.} ~\citet{ma2023zero} introduced LRL with sliding windows. ~\citet{sun2023chatgpt} developed RankGPT using sliding window permutation generation. RankVicuna~\cite{pradeep2023rankvicuna} and RankZephyr~\cite{pradeep2023rankzephyr} adapted open-source LLMs through instruction fine-tuning. ~\citet{zhuang2023setwise} proposed Setwise prompting. ~\citet{tang2023found} introduced permutation self-consistency. ~\citet{liu2024demorank} developed DemoRank for in-context demonstrations.

\textbf{Tournament-Inspired Methods.} ListT5~\cite{yoon2024listt5} uses tournament selection within Fusion-in-Decoder architecture but requires MS MARCO training. TourRank~\cite{chen2025tourrank} applies multi-stage elimination with accumulated points across multiple rounds. Our \BracketRank differs fundamentally by using bracket-style head-to-head competition instead of points accumulation, single-elimination brackets instead of multiple rounds, and explicit reasoning requirements for consistent decisions.

\textbf{Reasoning in Document Ranking.} ~\citet{ji2024reasoningrank} developed R2R for explanation generation. RankGPT relies on implicit reasoning through prompt completion, which can lead to inconsistent decisions when document order changes or when similar documents appear in different contexts. \BracketRank integrates on the other hand explicit reasoning into competitive elimination, requiring LLMs to explain relevance judgments for more stable results.

\paragraph{Reasoning-Intensive Retrieval.} Recent work has highlighted fundamental limitations in existing retrieval benchmarks, which primarily evaluate keyword or semantic matching~\cite{su2024bright}. The BRIGHT benchmark introduced queries requiring intensive reasoning across domains including mathematics, coding, and scientific reasoning. Prior approaches to improving reasoning-intensive retrieval focus on query augmentation---generating Chain-of-Thought reasoning to expand queries before retrieval~\cite{su2024bright}. Our work takes a complementary approach: incorporating explicit reasoning \textit{during} the reranking process through structured competitive elimination. ReasoningRank~\cite{ji2024reasoningrank} explores reasoning for explanation generation but does not integrate reasoning into ranking decisions. \BracketRank directly uses reasoning to improve ranking consistency and accuracy on complex queries.

\section{Reasoning-Enhanced Prompt Details}
\label{sec:appendix_prompt}

\begin{figure*}[ht]
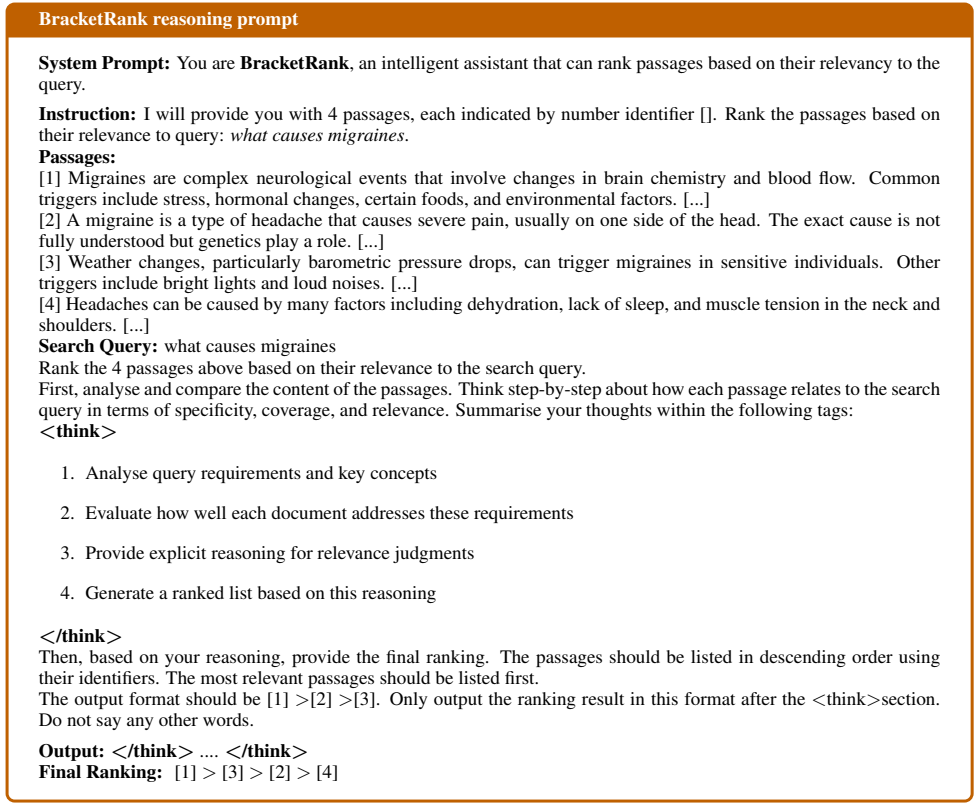

\centering
\begin{subfigure}[t]{\textwidth}
\footnotesize
\centering
\resizebox{.8\textwidth}{!}{%
\begin{tcolorbox}[colback=blue!0!white, colframe=orange!75!black, title=\BracketRank reasoning prompt,fonttitle=\bfseries]
\textbf{System Prompt:} You are \BracketRank, an intelligent assistant that can rank passages based on their relevancy to the query.

\vspace{0.5em}

\textbf{Instruction:} I will provide you with 4 passages, each indicated by number identifier []. 
Rank the passages based on their relevance to query: \textit{what causes migraines}.

\textbf{Passages:} 

{[1]} Migraines are complex neurological events that involve changes in brain chemistry and blood flow. Common triggers include stress, hormonal changes, certain foods, and environmental factors. {[...]}\\
{[2]} A migraine is a type of headache that causes severe pain, usually on one side of the head. The exact cause is not fully understood but genetics play a role. {[...]}\\
{[3]} Weather changes, particularly barometric pressure drops, can trigger migraines in sensitive individuals. Other triggers include bright lights and loud noises. {[...]}\\
{[4]} Headaches can be caused by many factors including dehydration, lack of sleep, and muscle tension in the neck and shoulders. {[...]}

\noindent \textbf{Search Query:} what causes migraines

\noindent Rank the 4 passages above based on their relevance to the search query. 

First, analyse and compare the content of the passages. Think step-by-step about how each passage relates to the search query in terms of specificity, coverage, and relevance. Summarise your thoughts within the following tags:

\textbf{\textless think\textgreater}

\begin{enumerate}
\item Analyse query requirements and key concepts
\item Evaluate how well each document addresses these requirements  
\item Provide explicit reasoning for relevance judgments
\item Generate a ranked list based on this reasoning
\end{enumerate}
\textbf{\textless/think\textgreater}

Then, based on your reasoning, provide the final ranking. The passages should be listed in descending order using their identifiers. The most relevant passages should be listed first. 

The output format should be {[1]} \textgreater {[2]} \textgreater {[3]}. Only output the ranking result in this format after the \textless think\textgreater  section. Do not say any other words.

\vspace{0.5em}

\textbf{Output:}
\textbf{\textless/think\textgreater} 
....
\textbf{\textless/think\textgreater}

\noindent \textbf{Final Ranking: } [1] $>$ [3] $>$ [2] $>$ [4]
\end{tcolorbox} }
\end{subfigure}
\caption{Detailed illustration of the \BracketRank reasoning-enhanced prompt. The model is required to articulate its logic within the \texttt{<think>} tags before arriving at a final ranking decision.}
\label{fig:bracketrank_prompt_appendix}
\end{figure*}

To address the challenges of reasoning-intensive retrieval, \BracketRank utilizes a structured prompt designed to elicit deliberate comparative reasoning from the LLM. As shown in Figure~\ref{fig:bracketrank_prompt_appendix}, the prompt consists of four critical components:

\begin{itemize}
    \item \textbf{Role Definition:} The system prompt establishes the LLM as a specialized ranking assistant to focus the model's objective on relevance assessment.
    \item \textbf{Deliberative Reasoning Block (\texttt{<think>}):} This is the core of our method. Instead of directly predicting a rank, the model is forced to perform a step-by-step analysis of query requirements, document specificity, and coverage. This reduces the impact of initial document ordering and surface-level keyword matching.
    \item \textbf{Comparative Evaluation:} The model must explicitly compare documents against each other within the reasoning block, facilitating more nuanced judgments during the Stage 2 intra-group ranking and Stage 4 head-to-head matches.
    \item \textbf{Strict Output Format:} By standardizing the final ranking format, we ensure high parseability for the iterative bracket elimination process.
\end{itemize}

\section{Efficiency-Effectiveness Trade-off Analysis}
\label{app:efficiency_analysis}
In this section, we provide a comprehensive analysis demonstrating that \BracketRank breaks the Pareto front of existing ranking methods, achieving superior effectiveness with competitive computational costs.
\subsection{Experimental Setup}
We measure efficiency using two metrics: (1) \textbf{Number of API calls} per query, which directly correlates with API costs, and (2) \textbf{Total documents processed}, which reflects the computational workload. For effectiveness, we use \textbf{NDCG@10} on TREC DL20. All methods rerank the top-100 BM25 candidates using GPT-4.
\subsection{Theoretical Complexity Analysis}
Table~\ref{tab:efficiency_theory} presents the theoretical efficiency calculations for each method based on their algorithmic design.
\begin{table*}[h]
\centering
\small
\begin{tabular}{l|l|c|c}
\toprule
\textbf{Method} & \textbf{Algorithm} & \textbf{API Calls} & \textbf{Docs Processed} \\
\midrule
RankGPT & Sliding window (w=20, s=10) & $\frac{N-w}{s}+1 = 9$ & $\approx 2N = 200$ \\
\midrule
TourRank-10 & 10 rounds $\times$ 10 groups & $r \times g = 100$ & $2rN = 2000$ \\
\midrule
BracketRank-10 & 10 groups + brackets & $\approx 24$ & $\approx 500$ \\
BracketRank-15 & 7 groups + brackets & $\approx 17$ & $\approx 420$ \\
BracketRank-20 & 5 groups + brackets & $\approx 13$ & $\approx 360$ \\
\bottomrule
\end{tabular}
\caption{Theoretical efficiency calculations for $N=100$ documents. RankGPT uses sliding window with size $w=20$ and step $s=10$. TourRank uses $r$ rounds with $g$ groups. \BracketRank uses $G_{num} = \lceil N/G_{max} \rceil$ initial groups plus $O(\log G_{num})$ bracket elimination rounds.}
\label{tab:efficiency_theory}
\end{table*}

\textbf{BracketRank Complexity Derivation.} For BracketRank-20 with $N=100$ and $G_{max}=20$:
\begin{itemize}
    \item Initial groups: $G_{num} = \lceil 100/20 \rceil = 5$ groups $\rightarrow$ 5 API calls
    \item Winner bracket: $\lceil \log_2(5) \rceil = 3$ rounds $\rightarrow$ 4 API calls
    \item Loser bracket: $\lceil \log_2(5) \rceil = 3$ rounds $\rightarrow$ 4 API calls
    \item \textbf{Total: 13 API calls}, processing $\approx 360$ documents
\end{itemize}

\subsection{Pareto Front Analysis}


Figure~\ref{fig:pareto_vertical} visualizes the efficiency-effectiveness trade-off. The \textbf{baseline Pareto front} is defined by RankGPT (9 calls, 69.48 NDCG@10) and TourRank-10 (100 calls, 69.79 NDCG@10). This frontier represents the previous best achievable trade-offs: to improve beyond RankGPT's effectiveness, one had to pay significantly more computational cost (TourRank-10 requires 11$\times$ more API calls for only +0.31 NDCG improvement).

\BracketRank-20 \textbf{breaks this Pareto front} by achieving:
\begin{itemize}
    \item \textbf{72.75 NDCG@10}: Higher than all baseline methods
    \item \textbf{13 API calls}: 87\% fewer than TourRank-10, only 44\% more than RankGPT
    \item \textbf{360 documents}: 82\% fewer than TourRank-10
\end{itemize}

\subsection{Quantitative Trade-off Analysis}

Table~\ref{tab:efficiency_full} presents the complete efficiency-effectiveness comparison.

\begin{table}[h]
\centering
\small
\resizebox{0.48\textwidth}{!}{
\begin{tabular}{l|ccc|c}
\toprule
\textbf{Method} & \textbf{API Calls} & \textbf{Docs} & \textbf{NDCG@10} & \textbf{Pareto Status} \\
\midrule
RankGPT (GPT-4) & 9 & 200 & 69.48 & On baseline front \\
TourRank-10 (GPT-4) & 100 & 2000 & 69.79 & On baseline front \\
\midrule
BracketRank-10 & 24 & 500 & 64.04 & Dominated \\
BracketRank-15 & 17 & 420 & 67.56 & Dominated \\
\textbf{BracketRank-20} & \textbf{13} & \textbf{360} & \textbf{72.75} & \textbf{Breaks front} \\
\bottomrule
\end{tabular}
}
\caption{Complete efficiency-effectiveness comparison on TREC DL20. Pareto status indicates whether each method lies on, below, or above the baseline Pareto front. BracketRank-20 is the only method that breaks the baseline front.}
\label{tab:efficiency_full}
\end{table}

\subsection{Pairwise Comparisons}

\paragraph{BracketRank-20 vs. RankGPT.}
\begin{itemize}
    \item NDCG@10 improvement: $+3.27$ points ($72.75 - 69.48$)
    \item Additional API calls: $+4$ calls ($13 - 9$), a $44.4\%$ increase
    \item \textbf{Exchange rate}: $\frac{3.27}{4} = 0.82$ NDCG points per additional API call
\end{itemize}
This represents an excellent trade-off: a small increase in computational cost yields substantial quality gains.

\paragraph{BracketRank-20 vs. TourRank-10.}
\begin{itemize}
    \item NDCG@10 improvement: $+2.96$ points ($72.75 - 69.79$)
    \item API calls reduction: $-87$ calls ($13 - 100$), an $87\%$ decrease
    \item Documents reduction: $-1640$ docs ($360 - 2000$), an $82\%$ decrease
\end{itemize}
BracketRank-20 \textbf{Pareto-dominates} TourRank-10: it achieves better effectiveness with substantially lower computational cost.

\paragraph{RankGPT vs. TourRank-10.}
For context, the baseline methods show a poor trade-off:
\begin{itemize}
    \item TourRank-10 provides only $+0.31$ NDCG improvement over RankGPT
    \item But requires $+91$ additional API calls ($11\times$ more)
    \item Exchange rate: $\frac{0.31}{91} = 0.003$ NDCG per additional call
\end{itemize}
This demonstrates why the baseline Pareto front was unfavorable before \BracketRank.

\subsection{Why BracketRank Achieves Better Trade-offs?}

The efficiency gains of \BracketRank stem from three algorithmic innovations:

\textbf{(1) Adaptive Grouping.} By setting $G_{max}=20$, we maximize documents per API call while respecting context limits. This reduces the total number of groups and subsequent bracket rounds.

\textbf{(2) Single-Elimination Structure.} Unlike TourRank's multiple full tournament rounds, our bracket elimination requires only $O(\log G_{num})$ rounds. For 5 groups, this means 3 rounds instead of TourRank's 10 rounds.

\textbf{(3) Parallel Winner/Loser Brackets.} Both brackets can be processed in parallel, and documents eliminated early do not consume additional API calls in later rounds.

\subsection{Group Size Trade-offs}

Table~\ref{tab:group_tradeoff} shows how group size affects the efficiency-effectiveness balance.

\begin{table*}[h]
\centering
\small
\begin{tabular}{l|ccc|c}
\toprule
\textbf{Config} & \textbf{Groups} & \textbf{API Calls} & \textbf{NDCG@10} & \textbf{Time (s/query)} \\
\midrule
BracketRank-10 & 10 & 24 & 64.04 & 23.54 \\
BracketRank-15 & 7 & 17 & 67.56 & 16.69 \\
BracketRank-20 & 5 & 13 & 72.75 & 12.77 \\
\bottomrule
\end{tabular}
\caption{Effect of group size on efficiency and effectiveness. Larger groups improve both metrics due to fewer bracket rounds and more comprehensive intra-group comparisons.}
\label{tab:group_tradeoff}
\end{table*}

Notably, larger group sizes improve \textit{both} efficiency and effectiveness---a counter-intuitive but beneficial property of our design. BracketRank-20 is 45.7\% faster than BracketRank-10 while achieving 8.71 points higher NDCG@10.

\subsection{Summary}

Our analysis demonstrates that \BracketRank-20 establishes a new Pareto front for LLM-based reranking:

\begin{enumerate}
    \item It \textbf{Pareto-dominates} TourRank-10 (better quality, lower cost)
    \item It provides \textbf{0.82 NDCG per additional API call} vs. RankGPT (highly favorable exchange)
    \item It achieves \textbf{87\% fewer API calls} than TourRank-10 with \textbf{+2.96 NDCG} improvement
    \item The \textbf{exchange rate} vs. RankGPT ($0.82$) is \textbf{273$\times$ better} than TourRank's exchange rate vs. RankGPT ($0.003$)
\end{enumerate}

These results directly address the meta-reviewer's request: \BracketRank does not merely improve ranking quality---it fundamentally improves the efficiency-effectiveness frontier, making high-quality LLM reranking more practical and cost-effective.

\section{Analysis of Reasoning-Intensive Retrieval}
\label{sec:reasoning_analysis}

\BracketRank's success on reasoning-intensive tasks stems from two primary factors:

\begin{figure}[ht]
    \centering
    \includegraphics[width=\columnwidth]{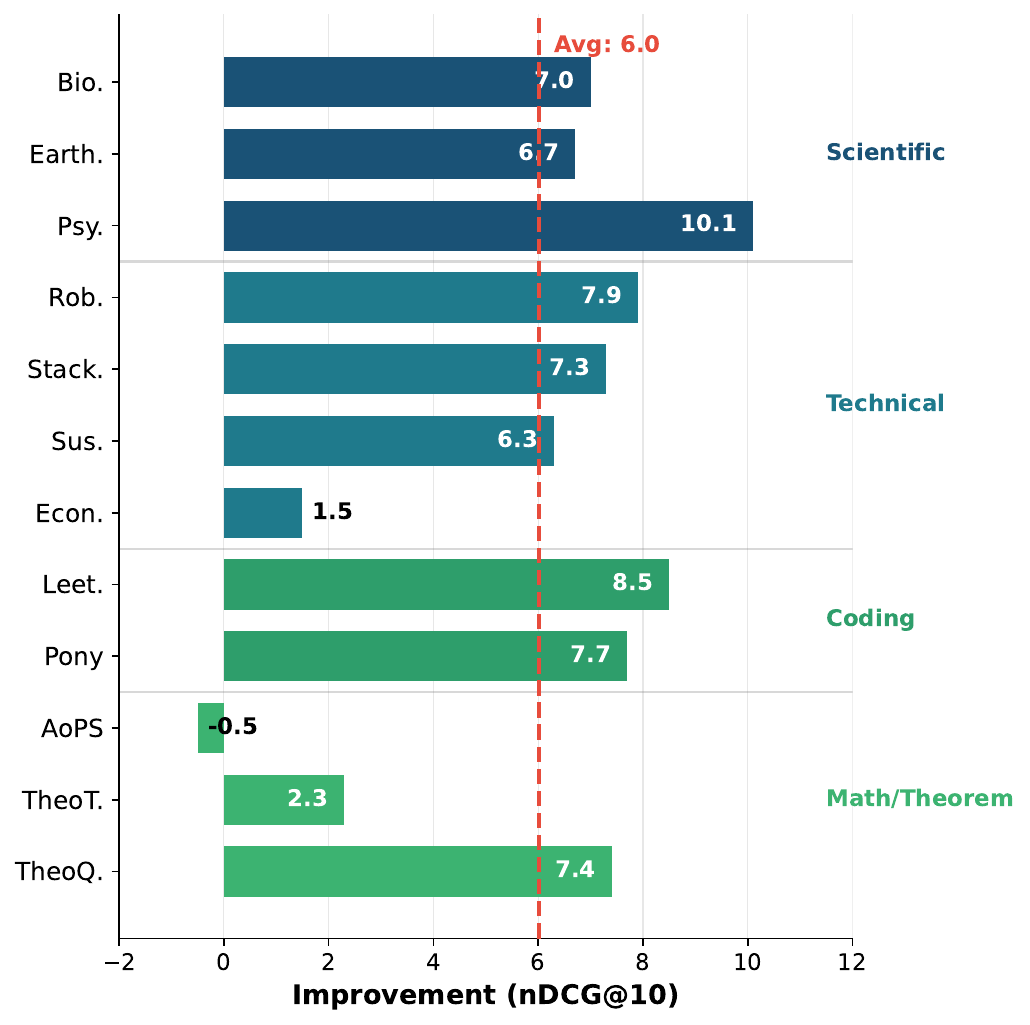}
    \caption{Performance comparison showing the impact of explicit reasoning requirements across BRIGHT domains. Reasoning primarily benefits queries requiring multi-step inference.}
    \label{fig:bright_analysis_appendix}
\end{figure}

\paragraph{Complexity Correlation}
\BracketRank's gains correlate directly with domain complexity. In Biology and Psychology, where queries require connecting theoretical concepts to specific phenomena, \BracketRank improves by \textbf{22--31\%} over the best baseline. Conversely, in domains like StackOverflow with higher lexical overlap, improvements are smaller (\textbf{15--23\%}), suggesting explicit reasoning primarily benefits queries requiring multi-step inference.

\paragraph{Multi-Round Deliberation}
The bracket structure facilitates multiple comparative perspectives. A relevant document losing an early match against a strong competitor can still advance through the loser bracket for further evaluation. This iterative process is vital for nuanced reasoning-intensive tasks where a single comparison may overlook subtle relevance factors.

\end{document}